# Integrating Macrostate Probability Distributions with Swing Adsorption Modeling for Binary/Ternary Gas Separation

Sunghyun Yoon[1,¶], Jui Tu[2,¶], Li-Chiang Lin[2,3,*], and Yongchul G. Chung[1,4,*]

[1]*School of Chemical Engineering, Pusan National University, Busan, 46241, Republic of Korea*

[2]*Department of Chemical Engineering, National Taiwan University, Taipei 106319, Taiwan*

[3]*William G. Lowrie Department of Chemical and Biomolecular Engineering, The Ohio State University, Ohio 43210, USA*

[4]*Graduate School of Data Science, Pusan National University, Busan, 46241, Republic of Korea*

[¶] *These authors contribute equally*

*Corresponding authors.

E-mail: lclin@ntu.edu.tw (L.-C. Lin), drygchung@gmail.com (Y.G. Chung)

## ABSTRACT

Accurate and efficient prediction of multicomponent adsorption equilibria across pressures, temperatures, and compositions remain a central challenge for designing energy-efficient adsorption-based separation processes. Traditional approaches, including model fitting and ideal adsorbed solution theory (IAST), often fail to balance accuracy, computational efficiency, and transferability under process-relevant conditions. Here, we introduce a material-to-process modeling framework that integrates macrostate probability distributions (MPDs) from flat-histogram Monte Carlo simulations with rigorous cyclic process optimization. MPDs directly capture the joint occupancy distributions of adsorbates, producing reweightable landscape that enables high-fidelity mixture adsorption equilibria without repeated simulations or model assumptions. We show that coupling this statistical mechanical foundation with process modeling delivers accurate and computationally efficient evaluations for binary and ternary gas mixture separations. This integration establishes MPD-based modeling as a generalized method for predictive multicomponent adsorption equilibria, accelerating the discovery and design of adsorbent materials for carbon capture and other separation challenges.

## 1. INTRODUCTION

Accurate and efficient prediction of multicomponent adsorption equilibria across arbitrary temperature, pressure, and composition is central to the design of energy-efficient adsorption-based separation processes in energy, environmental, and chemical manufacturing applications[1-6]. For a fixed temperature, pressure, and composition, such predictions are straightforward using direct experimentations or molecular simulations, such as grand canonical Monte Carlo (GCMC). However, chemical processes, such as cyclic adsorption processes, rarely operate on a single state point but over broad ranges of temperatures, pressures, and feed compositions. As such, the process optimizations typically require accurate equilibrium data spanning the entire operating envelope. Generating high-fidelity data is a formidable challenge because each new state point (T, P, and composition) typically demands a separate experiment or molecular simulation, making exhaustive mapping of multicomponent isotherms experimentally and computationally prohibitive.

Classical model-based approaches, such as the dual-site Langmuir or other fitted adsorption models, require parameter estimation that may not be robust outside the fitted conditions. The widely used ideal adsorbed solution theory (IAST), first proposed by Myers and Prausnitz in 1965, provides a model-free framework to predict mixture adsorption from pure isotherms and has been applied broadly in molecular simulation and materials screening campaigns[7-12]. However, for the former, Farmahini et al. demonstrated that applying different fitting procedures resulted in different parameter sets, which in turn resulted in variations in the predicted mixture adsorption isotherms[13]. These discrepancies ultimately led to deviations in process performance of up to 30%, highlighting the sensitivity of such models to the fitting method used. For the latter, while IAST can offer accurate predictions of mixture adsorption under certain conditions, its implicit formulation makes it computationally expensive when integrated into process modeling and optimization frameworks[14, 15]. Furthermore, violations of its key assumptions such as equal access to the adsorbent surface for all components and ideal solution behavior in the adsorbed phase can lead to substantial errors in the predicted mixture isotherms[16-23]. Mixture adsorption isotherms obtained directly from GCMC simulations are often regarded as ground truth and have been widely used for benchmarking[13, 15, 24, 25]. While direct GCMC simulations at each operating point remain the most rigorous option, they become intractable when hundreds or thousands of high-fidelity points are needed for process-level optimization.

Flat-histogram Monte Carlo methods, originally developed in statistical mechanics to uniformly sample each macrostate, provide a powerful alternative[26-29]. In particular, a recently developed 2D-NVT+W simulation approach[30], a variant of flat histogram Monte Carlo methods resembling the transition-matrix Monte Carlo (TMMC) scheme[27], offers a promising solution. In contrast to Wang-Landau sampling[26] that employs dynamically updating biasing functions, this method directly samples each macrostate. This approach determines the macrostate probability distribution (MPD) that represents the relative probabilities between each possible macrostate. Importantly, this distribution can be analytically reweighted to predict the mixture adsorption equilibrium under arbitrary conditions without the need for repetitive simulations or model fitting. Such a strategy is conceptually related to the Ferrenberg-Swendsen single- and multi-histogram methods[31-33], which perform canonical simulations at several parameter points, construct the density of states from the resulting energy histograms, and then use histogram reweighting to obtain thermodynamic properties at other conditions. The 2D-NVT+W approach has also been shown to reproduce GCMC-computed mixture adsorption isotherms, ensuring high accuracy[30]. While flat-histogram methods are established in molecular modeling of adsorption, they have not been integrated into process-level modeling of adsorption cycles.

In this work, we present a material-to-process modeling framework that couples MPD obtained from NVT+W simulations with rigorous process modeling and optimization. This coupling bridges molecular-scale thermodynamics and process-scale performance metrics, creating a standard workflow where a set of high-fidelity molecular simulation run can drive the full-scale process modeling and optimization. As a model system, we investigate the removal of acid gases from natural gas using zeolites, a process that is critical for improving fuel quality and preventing corrosion in downstream equipment[34]. We first screened a database of all-silica zeolites using adsorption energy distribution analysis to select two representative cases: one in which IAST provides accurate predictions and another where it fails to capture mixture adsorption behavior. For the selected zeolites, GCMC simulations were performed to generate pure-component adsorption isotherms and the NVT+W simulations were conducted to compute the corresponding MPDs of mixtures. The resulting isotherms were also fitted to the dual-site Langmuir-Freundlich (DSLF) model, and the obtained parameters were then used in both the extended dual-site Langmuir-Freundlich (EDSLF) and IAST frameworks to predict the mixture adsorption equilibrium. All three prediction methods, MPD-based, EDSLF-based, and IAST-based, were subsequently integrated into process models and their performances were compared in terms of prediction accuracy and computational efficiency. Starting with binary system ($CO_2$/$CH_4$), we also extended the analysis to ternary system ($H_2S$/$CO_2$/$CH_4$) to assess the generalizability and robustness of the MPD-based approach. Our results highlight the advantages of the integration of MPD framework in process modeling over classical methods and establish standardized workflow that can aid in the multi-scale adsorbent materials discovery campaign.

## 2. METHODS

### 2.1 Zeolite structures

In this study, all-silica zeolites were selected as candidate materials for natural gas upgrading, as in previous studies[15, 35], due to their inherently hydrophobic nature, which allows for the efficient removal of acidic gases even in the presence of water. The all-silica zeolite structures were obtained from International Zeolite Association (IZA) SC dataset (http://www.iza-structure.org/databases/)[36]. The dataset includes both the idealized framework structure in its all-silica form (designated as XYZ-0) and experimentally determined structures that incorporate non-silicon atoms at tetrahedral sites, labeled as XYZ-n (n = 1–6), comprising a total of 401 zeolite structures.

### 2.2 Molecular simulations

In all the Monte Carlo simulations conducted herein, intermolecular interactions were modeled as the sum of Coulombic and van der Waals (vdW) contributions. Coulombic interactions were computed using the Ewald summation method[37] with a relative error of $10^{-6}$. The vdW interactions were described by the 12-6 Lennard-Jones (LJ) potential, which was truncated at a cutoff distance of 10 Å with analytic tail corrections applied. The simulation cell dimensions were set to at least twice the cutoff radius in all directions. LJ parameters and partial charges for framework atoms were assigned based on the TraPPE-zeo force field[38]. As in previous studies, aluminum and phosphorus atoms were modeled using the same LJ parameters and partial charges as silicon atoms[15, 35]. $H_2S$, $CO_2$ and $CH_4$ molecules were represented using the TraPPE force field[39-41], with $CO_2$ modeled as a three-site molecule and $CH_4$ treated as a united-atom model, and $H_2S$ modeled as a four-site molecule, with corresponding LJ parameters and partial charges. All force field parameters used in this study are listed in Table S1. LJ parameters for unlike atom pairs were determined using the Lorentz–Berthelot combining rules.

GCMC simulations were carried out to predict the adsorption behavior of $H_2S$, $CO_2$ and $CH_4$ in zeolite frameworks, considering both single-component systems and their binary and ternary mixtures. Each simulation consisted of 300,000 cycles for initialization, followed by 300,000 production cycles for ensemble averages. For single-component isotherms, Monte Carlo (MC) moves including swap (insertion and deletion), translation, rotation, and reinsertion were employed with equal probabilities for sampling. For the mixture simulations, the identity change move was also included. Widom's particle insertion simulations were also performed with 20,000 cycles to calculate the isosteric heat of adsorption for $H_2S$, $CO_2$ and $CH_4$ in the zeolites at 298 K. The adsorption energy distributions for each zeolite were generated by inserting a single adsorbate molecule inside the zeolite framework and sampling the adsorbate–framework interaction energies using MC moves. Each distribution was constructed from 5,000 cycles, during which translation, rotation, and reinsertion moves were employed with equal probabilities. 2D and 3D NVT+W simulations were performed to compute the MPD of the binary and ternary mixtures, respectively. The simulations were conducted at a reference condition of 300 K and 0.1 bar pressure for each adsorbate (i.e., a total pressure of 0.2 bar for binary and 0.3 bar for ternary mixtures). Each NVT+W simulation was composed of 100,000 initialization cycles and an additional 100,000 production cycles, during which particle moves including translation, rotation, reinsertion, and insertion/deletion, were attempted with equal probabilities (1:1:1:1). Note though, in NVT+W simulations, insertion/deletion moves will never be accepted. During all simulations, the zeolite framework was treated as rigid. All simulations were performed using the open-source RASPA 2.0[42], with in-house modifications applied for the NVT+W simulations.

### 2.3 NVT+W method - theoretical formulation

The core of the NVT+W method first reported by Smit and co-workers[29] lies in its ability in determining the MPD by uniformly sampling each macrostate (i.e., $N$, the number of molecules in the adsorbent) under a reference chemical potential ($\mu$) and temperature ($T$). Specifically, simulations in the canonical ensemble are conducted for each possible macrostate, with Widom ghost insertions and deletions performed on the fly. This procedure is essentially the TMMC scheme[27] with a bin size of one and represents a distinct difference from the Wang-Landau method[26], where a flat histogram is generated by dynamically updating a biasing function and using biased acceptance rules. The acceptance ratios of these moves are accumulated in the C-matrix to determine the transition probability ($P(N \rightarrow N_p)$), representing the probability of transitioning from the sampled macrostate $N$ to a neighboring state $N_p$. Per detailed balance, the MPD or $\Pi(N; \mu VT)$ is obtained via eq 1:

$$P(N \rightarrow N_p)\Pi(N; \mu, V, T) = P(N_p \rightarrow N)\Pi(N_p; \mu, V, T) \tag{1}$$

Once the MPD is obtained, the adsorption uptake at the reference condition can be computed as the probability-weighted average over all possible macrostates. Moreover, the MPD can be analytically reweighted to any other condition ($\mu', V, T'$), yielding gas loading under any pressure and temperature (i.e., complete isotherm). The detailed equations will be discussed in the following section.

The NVT+W approach has been validated in several recent studies for adsorption of pure components[29, 43-46], demonstrating its accuracy and computational efficiency for adsorption calculations. Moreover, it has been previously extended to sample binary mixtures or two-dimensional macrostates ($\boldsymbol{N} = (N_1, N_2)$)[30]. Although the simulation protocol remains the same, the derivation of the 2D MPD poses challenges due to inconsistencies in transition pathways. For example, the probability of reaching the macrostate (1,1) from (0,0) via (1,0) may differ from that via (0,1). To resolve these mismatches, a simplified, so-called local optimization approach was proposed in prior work[30] to estimate relative probabilities without optimizing over all macrostates.

Herein, to support the optimization of adsorption processes involving ternary gas mixtures, we further extended the NVT+W method to three dimensions. In the 3D NVT+W framework, macrostates ($\boldsymbol{N} = (N_1, N_2, N_3)$) are defined within a triangular pyramidal region bounded by the vertices (0,0,0), ($N_1^{max}$, 0,0), (0, $N_2^{max}$, 0), and (0,0, $N_3^{max}$). $N_i^{max}$ represents the maximum number of possible molecules per simulation supercell for component $i$, and is determined from GCMC simulations conducted at a high pressure of 100 bar and a low temperature of 270 K, representing near-saturation conditions. The computation follows a hierarchical approach in which the pure-component MPDs (i.e., $\Pi(N_1, 0,0)$, $\Pi(0, N_2, 0)$, and $\Pi(0,0, N_3)$) were first determined. These were then used to construct 2D MPDs. Finally, the full 3D MPD was obtained by applying an extended local optimization procedure in three dimensions, using a generalized error function described in eq 2.

$$\Delta_{local}^2 = \sqrt{C(N_1 \to N_1-1; N_2; N_3)C(N_1-1 \to N_1; N_2; N_3)} \times \left(ln\frac{P(N_1 \to N_1-1; N_2; N_3)\Pi(N_1, N_2, N_3; \boldsymbol{\mu}, V, T)}{P(N_1-1 \to N_1; N_2; N_3)\Pi(N_1-1, N_2, N_3; \boldsymbol{\mu}, V, T)}\right)^2$$
$$+\sqrt{C(N_1; N_2 \to N_2-1; N_3)C(N_1; N_2-1 \to N_2; N_3)} \times \left(ln\frac{P(N_1; N_2 \to N_2-1; N_3)\Pi(N_1, N_2, N_3; \boldsymbol{\mu}, V, T)}{P(N_1; N_2-1 \to N_2; N_3)\Pi(N_1, N_2-1, N_3; \boldsymbol{\mu}, V, T)}\right)^2$$
$$+\sqrt{C(N_1; N_2; N_3 \to N_3-1)C(N_1; N_2; N_3-1 \to N_3)} \times \left(ln\frac{P(N_1; N_2; N_3 \to N_3-1)\Pi(N_1, N_2, N_3; \boldsymbol{\mu}, V, T)}{P(N_1; N_2; N_3-1 \to N_3)\Pi(N_1, N_2, N_3-1; \boldsymbol{\mu}, V, T)}\right)^2 \quad (2)$$

Although 3D NVT+W enables accurate and rapid computation of gas uptake under any temperature, pressure, and composition, exhaustive sampling of all macrostates may be computationally prohibitive. To illustrate, in the case of zeolite GIS-1, the total number of macrostates exceeds 25,000, making the simulation infeasible and limiting its practical utility. To address this, we have adopted an equal-space sampling[30] strategy in which only macrostates separated by a fixed interval (e.g., $\Delta N = 4$) are sampled. Specifically, setting $\Delta N = 4$ means that only macrostates with $N = 0, 4, \ldots, N_{max}$ were sampled for each adsorbate. This leads to approximately a $4^3 = 64$-fold reduction in the total number of sampled macrostates in GIS, reducing the count from over 25,000 to only approximately 500 in GIS-1, while maintaining sufficient resolution for reweighting and uptake predictions.

## 2.4 Mixture adsorption equilibrium model

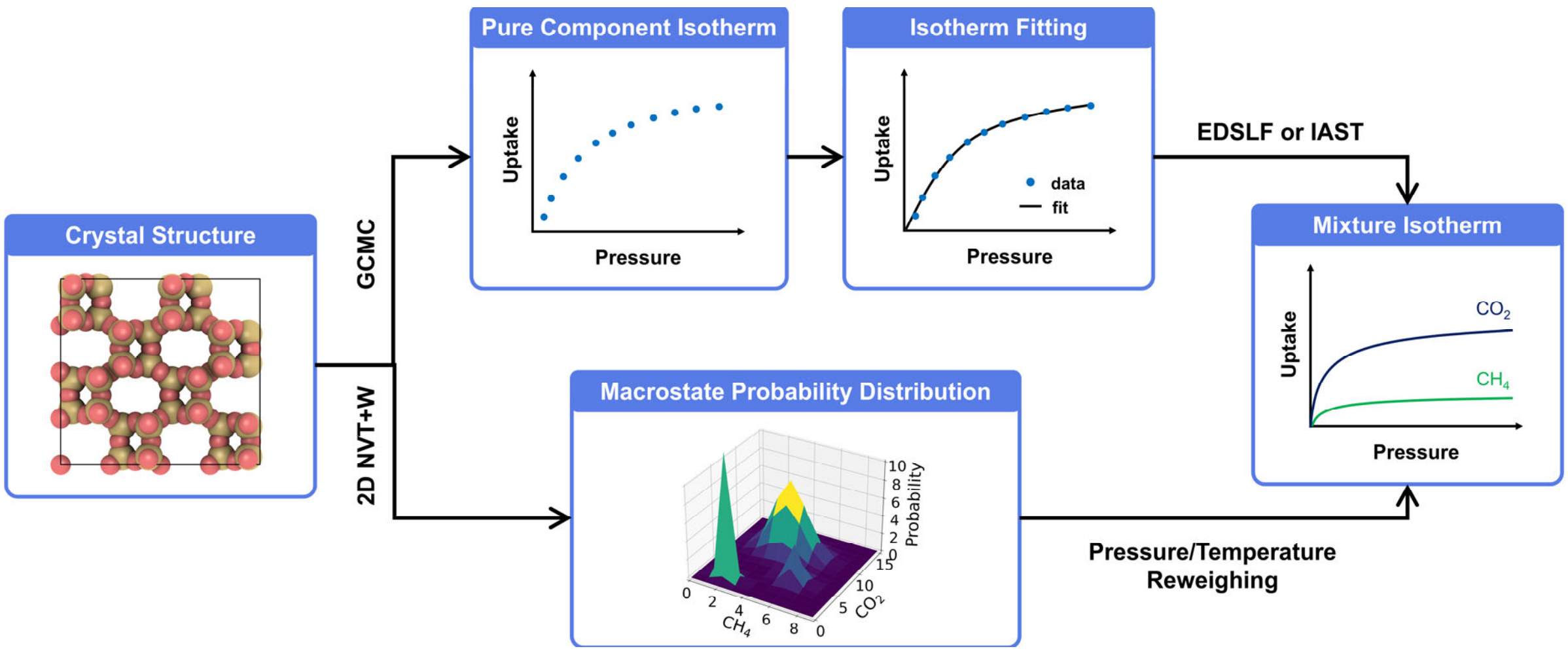

**Scheme 1.** Approaches for predicting mixture adsorption isotherms from crystal structures via molecular simulations.

As mentioned above, three different approaches were employed in this study to predict mixture adsorption equilibrium and were subsequently incorporated into the process model. The first two methods, the EDSLF model and the IAST, both rely on isotherm fitting of single-component adsorption data. Specifically, the EDSLF model uses only the fitted isotherm parameters, whereas the IAST approach requires both the fitted parameters and the DSLF model itself to predict mixture adsorption. The third approach utilizes the NVT+W method, wherein binary and ternary adsorption isotherms are respectively obtained by reweighting the 2D and 3D MPD obtained under a reference condition to other conditions (e.g., temperature, pressure, composition) and computing weighted averages. A schematic representation of the binary mixture isotherm prediction workflow for each approach is provided in Scheme 1, and the corresponding models are discussed in detail in the following subsections.

### 2.4.1 EDSLF model

Using binary mixture as an example, the single-component adsorption data for $CO_2$ and $CH_4$ obtained from GCMC simulations were fitted to the DSLF model[47] as shown in eq 3:

$$q_i^* = \frac{q_{sb,i} b_i P^{\frac{1}{n_{b,i}}}}{1 + b_i P^{\frac{1}{n_{b,i}}}} + \frac{q_{sd,i} d_i P^{\frac{1}{n_{d,i}}}}{1 + d_i P^{\frac{1}{n_{d,i}}}} \tag{3}$$

where $q_i^*$ represents the solid-phase equilibrium loading (mmol/g) of component $i$; $q_{sb,i}$ and $q_{sd,i}$ are saturation loadings (mmol/g) of component $i$ at sites 1 and 2, respectively; $b_i$ and $d_i$ are the affinity coefficients of component $i$ at sites 1 and 2, respectively; $n_{b,i}$ and $n_{d,i}$ are the ideal homogeneous surface deviations of component $i$ at sites 1 and 2, respectively; and $P$ is the gas-phase pressure (Pa). The EDSLF model with the Clausius–Clapeyron relationship[14, 48] was employed to predict mixture adsorption isotherms, as shown in eq 4:

$$q_{i,mix}^* = \frac{q_{sb,i} b_i \left(p_i e^{\frac{-\Delta H_i}{R}\left(\frac{1}{T} - \frac{1}{298}\right)}\right)^{\frac{1}{n_{b,i}}}}{1 + \sum_{i=1}^{n_{comp}} b_i \left(p_i e^{\frac{-\Delta H_i}{R}\left(\frac{1}{T} - \frac{1}{298}\right)}\right)^{\frac{1}{n_{b,i}}}} + \frac{q_{sd,i} d_i \left(p_i e^{\frac{-\Delta H_i}{R}\left(\frac{1}{T} - \frac{1}{298}\right)}\right)^{\frac{1}{n_{d,i}}}}{1 + \sum_{i=1}^{n_{comp}} d_i \left(p_i e^{\frac{-\Delta H_i}{R}\left(\frac{1}{T} - \frac{1}{298}\right)}\right)^{\frac{1}{n_{d,i}}}} \tag{4}$$

where $q_{i,mix}^*$ represents the solid-phase equilibrium loading (mmol/g) of component $i$ in the mixture; $p_i$ is the gas-phase partial pressure (Pa) of component $i$; $n_{comp}$ is the number of components in the mixture; $\Delta H_i$ denotes the isosteric heat of adsorption (kJ/mol) obtained from the Widom's particle insertion simulations; $R$ is the universal gas constant; and $T$ is the temperature (K). The DSLF parameters obtained from the fitting of single-component adsorption data were used as input for this model.

### 2.4.2 IAST method

The IAST, developed by Myers and Prausnitz[7], is a widely used thermodynamic framework for predicting multi-component adsorption equilibrium based solely on pure-component adsorption data[4]. The theory is based on several key assumptions: (1) the changes in the thermodynamic properties of the adsorbent upon gas adsorption are negligible compared to those of the adsorbate; (2) all adsorbate species have equal access to the same adsorption surface area; (3) the Gibbs dividing surface is used to define the adsorbed phase; and (4) the gas phase behaves as an ideal gas, while the adsorbed phase is treated as an ideal solution. Based on these assumptions, the set of equations governing IAST is derived as follows:

$$\pi = \pi_1(p_1^0) = \pi_2(p_2^0) = \cdots \tag{5}$$

$$\pi_i(p_i^0) = \frac{RT}{A} \int_0^{p_i^0} \frac{q_i^*(P)}{P} dP \tag{6}$$

$$p_i^0 = \frac{y_i P}{x_i} \tag{7}$$

$$\frac{1}{q_{T,mix}^*} = \sum_{i=1}^{n_{comp}} \frac{x_i}{q_i^*(p_i^0)} \tag{8}$$

where $\pi_i$ is the spreading pressure of component $i$; $p_i^0$ is the hypothetical sorption pressure of pure component $i$ that would yield the same spreading pressure as that of the mixture at equilibrium; $A$ is the surface area; $q_i^*$ is the adsorption isotherm of pure component $i$, which is obtained using the previously fitted parameters and the DSLF model; $y_i$ and $x_i$ are the gas phase and adsorbed phase mole fractions of component $i$, respectively; $q_{T,mix}^*$ is the total amount adsorbed (mmol/g) in the mixture. The non-linear system of equations (eqs 5–7) was solved using the *fsolve* function in MATLAB to obtain $x_i$ and $p_i^0$. These values were subsequently used in eq 8 to calculate $q_{T,mix}^*$, which allows for the prediction of the mixture adsorption isotherm.

### 2.4.3 MPD-based approach

The computed MPD can be analytically reweighted to any conditions. Specifically, using binary mixtures as illustration, to reweight the 2D MPD obtained at reference condition (i.e., $\boldsymbol{\mu}VT$) to a new condition (i.e., $\boldsymbol{\mu}'VT'$), eq 9 can be applied:

$$ln\frac{\Pi(\boldsymbol{N};\boldsymbol{\mu}',V,T')}{\Pi(\boldsymbol{0};\boldsymbol{\mu}',V,T')} = ln\frac{\Pi(\boldsymbol{N};\boldsymbol{\mu},V,T)}{\Pi(\boldsymbol{0};\boldsymbol{\mu},V,T)} + N_1\,ln\frac{f_1'}{f_1} + N_2\,ln\frac{f_2'}{f_2} + (N_1+N_2)\,ln\frac{T}{T'} + \sum_{n=1}^{\infty}\frac{1}{n!}\frac{\partial^n lnQ_c(\boldsymbol{N},V,\beta)}{\partial\beta^n}(\beta'-\beta)^n \quad (9)$$

where $f_1$, $f_2$, and $T$ represent the fugacities of component 1 and component 2, and the temperature under the reference condition, respectively, while $f_1'$, $f_2'$, and $T'$ denote the corresponding fugacities and temperature under the new condition; $Q_c$ represents the configurational part of the canonical partition function and $\beta = 1/k_BT$. The second and third terms on the right-hand side of eq 9 correspond to pressure reweighting, while the fourth and fifth terms are associated with temperature reweighting. The detailed derivation of the expression can be found in previous literature[29, 30]. In this study, we set the fugacity coefficient to be 1, so that, strictly speaking, the partial pressure of each component is equal to its fugacity. Furthermore, based on prior findings that the first-order term of the Taylor expansion alone provides sufficiently accurate results and that the second-order term only marginally improves the accuracy[30], we consider only the first-order term. This term can be expressed as eq 10 through the ensemble average of internal energy ($\langle E\rangle$).

$$\frac{\partial lnQ_c(\boldsymbol{N},V,\beta)}{\partial\beta} = -\langle E\rangle \quad (10)$$

Taking all of these aspects into account, the equation for reweighting the 2D MPD from condition $\boldsymbol{\mu}VT$ to condition $\boldsymbol{\mu}'VT'$ is expressed as eq 11:

$$ln\frac{\Pi(\boldsymbol{N};\boldsymbol{\mu}',V,T')}{\Pi(\boldsymbol{0};\boldsymbol{\mu}',V,T')} = ln\frac{\Pi(\boldsymbol{N};\boldsymbol{\mu},V,T)}{\Pi(\boldsymbol{0};\boldsymbol{\mu},V,T)} + N_1\,ln\frac{p_1'}{p_1} + N_2\,ln\frac{p_2'}{p_2} + (N_1+N_2)\,ln\frac{T}{T'} - \langle E\rangle(\beta'-\beta) \quad (11)$$

where $p_1$ and $p_2$ represent the partial pressures of component 1 and component 2 under the reference condition, respectively, and $p_1'$ and $p_2'$ denote the corresponding partial pressures under the new condition. Subsequently, the average loading values under condition $\boldsymbol{\mu}'VT'$ can be computed as the MPD-weighted average number of molecules using eqs 12 and 13:

$$\langle N_1\rangle_{\boldsymbol{\mu}'VT'} = \frac{\sum_{i=0}^{N_{1,max}}\left(i\sum_{j=0}^{N_{2,max}}\frac{\Pi(\boldsymbol{N};\boldsymbol{\mu}',V,T')}{\Pi(\boldsymbol{0};\boldsymbol{\mu}',V,T')}\right)}{\sum_{i=0}^{N_{1,max}}\left(\sum_{j=0}^{N_{2,max}}\frac{\Pi(\boldsymbol{N};\boldsymbol{\mu}',V,T')}{\Pi(\boldsymbol{0};\boldsymbol{\mu}',V,T')}\right)} \tag{12}$$

$$\langle N_2\rangle_{\boldsymbol{\mu}'VT'} = \frac{\sum_{j=0}^{N_{2,max}}\left(j\sum_{i=0}^{N_{1,max}}\frac{\Pi(\boldsymbol{N};\boldsymbol{\mu}',V,T')}{\Pi(\boldsymbol{0};\boldsymbol{\mu}',V,T')}\right)}{\sum_{j=0}^{N_{2,max}}\left(\sum_{i=0}^{N_{1,max}}\frac{\Pi(\boldsymbol{N};\boldsymbol{\mu}',V,T')}{\Pi(\boldsymbol{0};\boldsymbol{\mu}',V,T')}\right)} \tag{13}$$

$\langle N_1\rangle_{\boldsymbol{\mu}'VT'}$ and $\langle N_2\rangle_{\boldsymbol{\mu}'VT'}$ represent the average loading values (molecule/unitcell) of components 1 and 2, respectively, and can be converted to $q^*_{1,mix}$ and $q^*_{2,mix}$ (in mmol/g) by multiplying appropriate unit conversion factors.

Similarly, the 3D MPD obtained at the reference condition (i.e., $\boldsymbol{\mu}VT$) can be reweighted to a new condition (i.e., $\boldsymbol{\mu}'VT'$) using the following expression (eq 14).

$$ln\frac{\Pi(\boldsymbol{N};\boldsymbol{\mu}',V,T')}{\Pi(\boldsymbol{0};\boldsymbol{\mu}',V,T')} = ln\frac{\Pi(\boldsymbol{N};\boldsymbol{\mu},V,T)}{\Pi(\boldsymbol{0};\boldsymbol{\mu},V,T)} + N_1\,ln\frac{p_1'}{p_1} + N_2\,ln\frac{p_2'}{p_2} + N_3\,ln\frac{p_3'}{p_3} + (N_1+N_2+N_3)\cdot ln\frac{T}{T'} - \langle E\rangle(\beta'-\beta) \tag{14}$$

where $p_3$ and $p_3'$ represent the partial pressures of component 3 at the reference and new conditions, respectively. The corresponding adsorption uptakes can then be calculated from eqs 15–17.

$$\langle N_1\rangle_{\boldsymbol{\mu}'VT'} = \frac{\sum_{i=0}^{N_{1,max}}\left(\sum_{j=0}^{N_{2,max}}\sum_{k=0}^{N_{3,max}} i\,\frac{\Pi(\boldsymbol{N};\boldsymbol{\mu}',V,T')}{\Pi(\boldsymbol{0};\boldsymbol{\mu}',V,T')}\right)}{\sum_{i=0}^{N_{1,max}}\left(\sum_{j=0}^{N_{2,max}}\sum_{k=0}^{N_{3,max}}\frac{\Pi(\boldsymbol{N};\boldsymbol{\mu}',V,T')}{\Pi(\boldsymbol{0};\boldsymbol{\mu}',V,T')}\right)} \tag{15}$$

$$\langle N_2\rangle_{\boldsymbol{\mu}'VT'} = \frac{\sum_{i=0}^{N_{1,max}}\left(\sum_{j=0}^{N_{2,max}}\sum_{k=0}^{N_{3,max}} j\,\frac{\Pi(\boldsymbol{N};\boldsymbol{\mu}',V,T')}{\Pi(\boldsymbol{0};\boldsymbol{\mu}',V,T')}\right)}{\sum_{i=0}^{N_{1,max}}\left(\sum_{j=0}^{N_{2,max}}\sum_{k=0}^{N_{3,max}}\frac{\Pi(\boldsymbol{N};\boldsymbol{\mu}',V,T')}{\Pi(\boldsymbol{0};\boldsymbol{\mu}',V,T')}\right)} \tag{16}$$

$$\langle N_3\rangle_{\boldsymbol{\mu}'VT'} = \frac{\sum_{i=0}^{N_{1,max}}\left(\sum_{j=0}^{N_{2,max}}\sum_{k=0}^{N_{3,max}} k\,\frac{\Pi(\boldsymbol{N};\boldsymbol{\mu}',V,T')}{\Pi(\boldsymbol{0};\boldsymbol{\mu}',V,T')}\right)}{\sum_{i=0}^{N_{1,max}}\left(\sum_{j=0}^{N_{2,max}}\sum_{k=0}^{N_{3,max}}\frac{\Pi(\boldsymbol{N};\boldsymbol{\mu}',V,T')}{\Pi(\boldsymbol{0};\boldsymbol{\mu}',V,T')}\right)} \tag{17}$$

### 2.5 Dynamic process modeling and optimization

#### 2.5.1 Details of pressure/vacuum swing adsorption (PVSA) cycle

In this study, a five-step modified Skarstrom cycle[49] was employed as a model cycle. As shown in Scheme 2, the cycle consists of five steps: pressurization (Pres), adsorption (Ads), heavy reflux (HR), depressurization (Depres), and light reflux (LR). The cycle begins with the pressurization step, during which the feed gas is introduced into the column from the bottom, increasing the column pressure from the desorption pressure ($P_L$) to the adsorption pressure ($P_H$). During the adsorption step, the feed gas continues to flow into the column, where the $CO_2$ is selectively adsorbed while the non-adsorbed $CH_4$ exits the column from the top. In the heavy reflux step, the feed stream is stopped, and the heavy product collected during the light reflux step is introduced into the column through the inlet. During this step, additional $CH_4$ is released from the top end of the column. In the subsequent depressurization step, the column pressure is dropped back to the desorption pressure ($P_L$), which enables $CO_2$ to be desorbed and exit from the bottom of the column. Finally, in the light reflux step, the light product collected during the adsorption step is used to purge the residual $CO_2$ from the column.

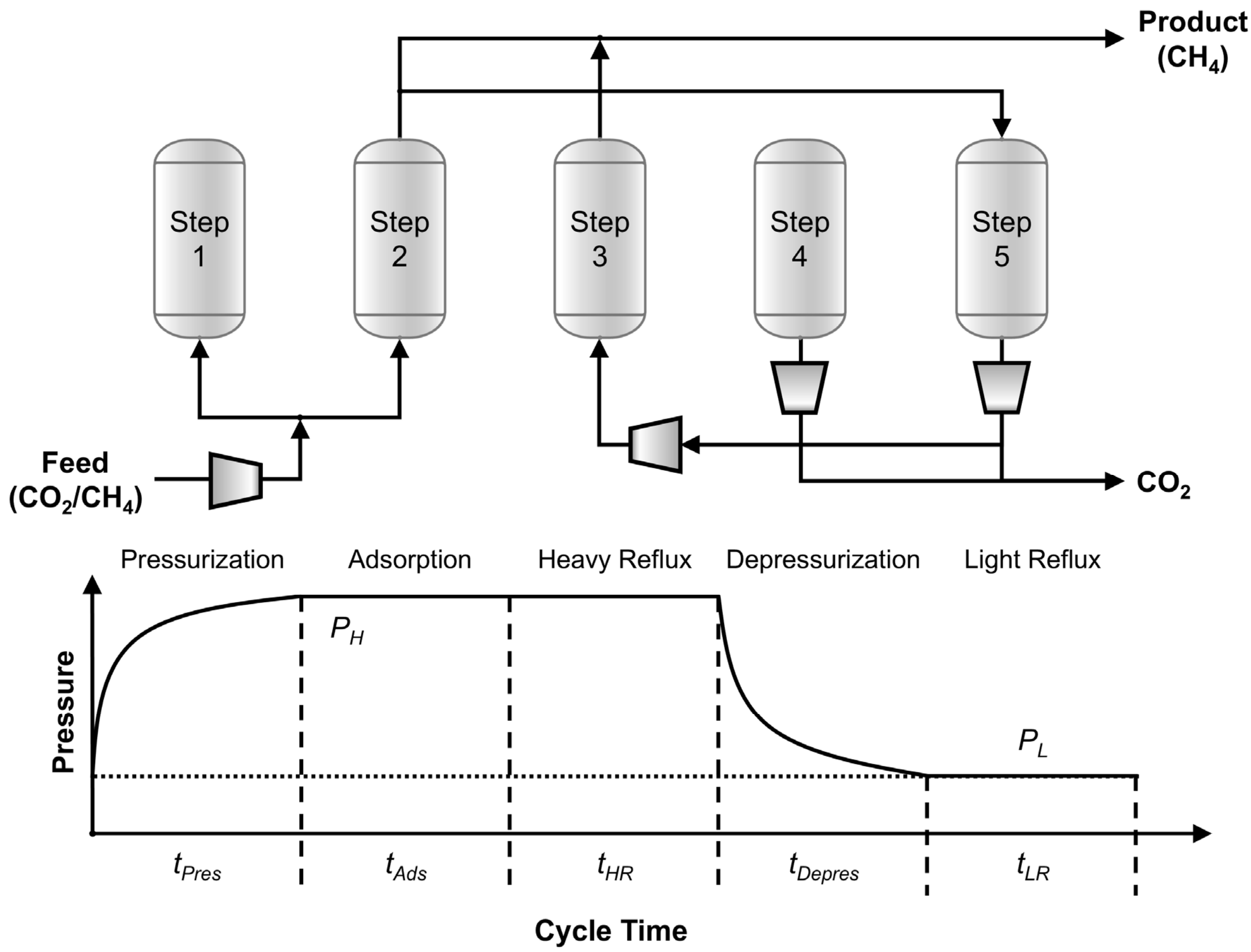

**Scheme 2.** Schematic representation of the five-step modified Skarstrom PVSA cycle. The cycle consists of the following steps: pressurization (*Pres*), adsorption (*Ads*), heavy reflux (*HR*), depressurization (*Depres*), and light reflux (*LR*). $P_H$ indicates the adsorption pressure, and $P_L$ indicates the desorption pressure.

**Table 1. All parameters used in the PVSA cycle simulation and optimization.**

| Parameter | Unit | Value | Type |
|---|---|---|---|
| ***Column properties*** | | | |
| Column length | m | 1.0 | Constant |
| Column diameter | m | 0.33 | Constant |
| Column void fraction | - | 0.40 | Constant |
| ***Adsorbent properties*** | | | |
| Pellet radius | m | $5.0\times10^{-3}$ | Constant |
| Adsorbent density | $kg/m^3$ | 1556.5 for AFG-1<br>1511.5 for GIS-1 | Constant |
| Specific heat capacity of adsorbent | J/kg/K | 750 | Constant |
| ***Gas properties*** | | | |
| Specific heat capacity of gas phase | J/mol/K | 35.80 | Constant |
| Specific heat capacity of adsorbed phase | J/mol/K | 35.80 | Constant |
| Fluid viscosity | kg/m/s | $1.13\times10^{-5}$ | Constant |
| Molecular diffusivity | $m^2/s$ | $1.30\times10^{-5}$ | Constant |
| Effective gas thermal conductivity | J/m/K/s | 0.09 | Constant |
| Mass transfer coefficient of $H_2S$ | 1/s | 0.18 | Constant |
| Mass transfer coefficient of $CO_2$ | 1/s | 0.16 | Constant |
| Mass transfer coefficient of $CH_4$ | 1/s | 0.20 | Constant |
| ***Scaling parameters*** | | | |
| $P_0$ | bar | Adsorption pressure | Variable |
| $T_0$ | K | Feed temperature | Constant |
| $v_0$ | m/s | Feed velocity | Variable |
| $q_{s,0}$ | mmol/g | 7.40 | Constant |
| ***Operating conditions*** | | | |
| Adsorption pressure | bar | [2.0, 10.0] | Variable |
| Feed velocity | m/s | [0.1, 0.8] | Variable |
| Desorption pressure | bar | [0.1, 2.0] | Variable |
| Light reflux ratio | - | [0.01, 0.99] | Variable |
| Heavy reflux ratio | - | [0.01, 0.99] | Variable |
| Adsorption time | s | [10, 500] | Variable |
| Heavy reflux time | s | Adsorption time | Variable |
| Light reflux time | s | Adsorption time | Variable |
| Depressurization time | s | [10, 50] | Variable |
| Pressurization time | s | [10, 50] | Variable |
| Feed pressure | bar | 1.0 | Constant |
| Feed temperature | K | 298.15 | Constant |

### 2.5.2 PVSA cycle model

A non-isothermal, one-dimensional mathematical model, developed by Leperi et al.[50] and Yancy-Caballero et al.[51], was employed and modified in this study to simulate the PVSA cycle. The model consists of a set of partial differential equations (PDEs) describing mass, energy, and momentum balances within the adsorption column, coupled with a linear driving force (LDF) model and the mixture adsorption equilibrium model. The detailed equations are provided in Table S2. The model is based on the following assumptions for the adsorption column: 1) the gas flow in the column is represented as an axially dispersed plug flow, 2) the gas phase is assumed to be ideal, 3) thermal equilibrium is assumed between the gas and solid phases, 4) the column void fraction and particle size remain constant throughout the column, and 5) radial gradients in concentration, temperature, and pressure are neglected. The system of PDEs was first transformed into a non-dimensional form for numerical stability and then discretized along the spatial direction of the column using a finite volume method[52] with a weighted essentially non-oscillatory (WENO) scheme[53]. A total of 30 finite volumes were used in spatial discretization. The resulting system of time-dependent ordinary differential equations (ODEs) was solved in MATLAB using the ode15s solver[54] with appropriate initial and boundary conditions. Each step of the PVSA cycle was modeled with specific boundary conditions, which are summarized in Table S3. A uni-bed approach was used to simulate the PVSA cycle, which was iterated until the system reached cyclic steady state (CSS). The system was considered to have reached CSS when the following two criteria were simultaneously satisfied: (1) the normalized state variables at the final condition of the last step of the cycle are within a tolerance of 0.01 of those at the initial condition of the first step, and (2) the total amount of gas leaving the column should be within 0.99 and 1.01 of the amount of gas entering the column. The maximum number of consecutive cycle iterations was set at 250. If the CSS was not achieved within this limit, the simulation was considered not converged and discarded. Once CSS was reached, $CH_4$ purity ($Pu_{CH_4}$) and recovery ($Re_{CH_4}$) were calculated using the eqs 18 and 19:

$$CH_4\ Purity, Pu_{CH_4} = \frac{n_{CH_4}^{Out\ from\ Ads} \times (1-\alpha_{LR}) + n_{CH_4}^{Out\ from\ HR}}{n_{total}^{Out\ from\ Ads} \times (1-\alpha_{LR}) + n_{total}^{Out\ from\ HR}} \quad (18)$$

$$CH_4\ Recovery, Re_{CH_4} = \frac{n_{CH_4}^{Out\ from\ Ads} \times (1-\alpha_{LR}) + n_{CH_4}^{Out\ from\ HR}}{n_{CH_4}^{In\ to\ Pres} + n_{CH_4}^{In\ to\ Ads}} \quad (19)$$

where $n_{CH_4}^{Out\ from\ Ads}$ and $n_{CH_4}^{Out\ from\ HR}$ represent the number of $CH_4$ moles in the outlet streams from the adsorption and heavy reflux steps, respectively; $n_{total}^{Out\ from\ Ads}$ and $n_{total}^{Out\ from\ HR}$ are the total numbers of moles in the outlet streams from the adsorption and heavy reflux steps, respectively; $n_{CH_4}^{In\ to\ Pres}$ and $n_{CH_4}^{In\ to\ Ads}$ denote the number of $CH_4$ moles in the inlet streams from the pressurization and adsorption steps, respectively; $\alpha_{LR}$ is the light reflux ratio. All parameters used for the PVSA cycle simulation are listed in Table 1.

### 2.5.3 Process cycle optimization

The PVSA cycle optimization was performed to simultaneously maximize $CH_4$ purity and recovery. To achieve this, the problem was formulated as a multi-objective optimization task, as described in eq 20:

$$minimize \quad J_1 = (1 - CH_4\ Purity)^2$$
$$minimize \quad J_2 = (1 - CH_4\ Recovery)^2$$
$$s.t. \quad CH_4\ Purity \geq y_{CH_4,0}$$
$$CH_4\ Recovery \geq 0.90 \tag{20}$$

where $y_{CH_4,0}$ is $CH_4$ mole fraction in the feed gas. To obtain the optimal cycle configuration, eight decision variables were considered: adsorption pressure ($P_H$), feed velocity ($v_F$), desorption pressure ($P_L$), light reflux ratio ($\alpha_{LR}$), heavy reflux ratio ($\beta_{HR}$), adsorption time ($t_{Ads}$), depressurization time ($t_{Depres}$), and pressurization time ($t_{Pres}$). The lower and upper bounds for these variables are summarized in Table 1. We solved the optimization problem using the non-dominated sorting genetic algorithm (NSGA-II)[55] implemented in MATLAB, with a population size of 80 and 80 generations. All optimizations were run in parallel on 20 Intel® Xeon® Gold 6448Y CPUs.

## 2.6 Techno-economic model

### 2.6.1 Design of PVSA system

A PVSA system was considered to treat natural gas feed. The PVSA system consists of $M$ parallel trains, each comprising $N_{col}$ adsorption columns, $N_{v,Depres}$ vacuum pumps for the depressurization step, $N_{v,LR}$ vacuum pumps for the light reflux step, one compressor for the feed, and one compressor for the heavy reflux step. A natural gas feed flow rate of 10,000 $m^3$(STP)/h was assumed for the analysis. The procedure proposed by Khurana and Farooq[56] was adopted to design the PVSA system.

The minimum number of adsorption columns ($N_{col}$) and vacuum pumps ($N_{v,j}$) required for each PVSA train to enable continuous operation were calculated using eqs 21 and 22, respectively:

$$N_{col} = ceiling(\frac{\sum_{i=steps} t_i}{t_{Pres} + t_{Ads}}) \tag{21}$$

$$N_{v,j} = ceiling\left(\frac{t_j}{t_{Pres} + t_{Ads}}\right) \ (j = Depres/LR) \tag{22}$$

where $t_i$ is the duration (s) of each step in the PVSA cycle. The number of parallel trains ($M$) was calculated using eq 23:

$$M = ceiling\left(\frac{\dot{n}_{CH_4,total}}{\dot{n}_{CH_4,in}}\right) \tag{23}$$

where $\dot{n}_{CH_4,total}$ is the total molar flow rate (mol $CH_4$/s) of $CH_4$ in the natural gas feed, and $\dot{n}_{CH_4,in}$ is the average molar flow rate (mol $CH_4$/s) of $CH_4$ fed into each train, defined by eq 24:

$$\dot{n}_{CH_4,in} = \frac{1}{t_{Pres} + t_{Ads}}\left(\int_0^{t_{Pres}} \dot{n}_{CH_4,Pres}\, dt + \int_0^{t_{Ads}} \dot{n}_{CH_4,Ads}\, dt\right) \tag{24}$$

where $\dot{n}_{CH_4,Pres}$ is the inlet molar flow rate (mol $CH_4$/s) of $CH_4$ during the pressurization step, and $\dot{n}_{CH_4,Ads}$ is the inlet molar flow rate (mol $CH_4$/s) of $CH_4$ during the adsorption step.

### 2.6.2 Cost model

The estimation of capital and operating costs was based on the methodologies and equations provided by Turton et al.[57]

#### 2.6.2.1 Capital cost

First, the costs of the PVSA system components, including adsorption columns, vacuum pumps, compressors, and compressor electric motors, were estimated. The purchase costs of the adsorption columns, vacuum pumps, compressors, and compressor electric motors were calculated using eqs 25, 26, 27, and 28, respectively:

$$log_{10} C_{P,col} = 3.4974 + 0.4485\, log_{10} V + 0.1074(log_{10} V)^2 \tag{25}$$

$$log_{10} C_{P,v} = 3.3892 + 0.0536\, log_{10} \dot{W}_v + 0.1536(log_{10} \dot{W}_v)^2 \tag{26}$$

$$log_{10} C_{P,comp} = 2.2897 + 1.3604\, log_{10} \eta\dot{W}_c - 0.1027(log_{10} \eta\dot{W}_c)^2 \tag{27}$$

$$log_{10} C_{P,drive} = 1.956 + 1.7142\, log_{10} \dot{W}_c - 0.2282(log_{10} \dot{W}_c)^2 \tag{28}$$

where $C_{P,col}$ is the purchase cost (\$) of adsorption column, $V = \pi {r_{in}}^2 L$ is the volume of the adsorption column ($m^3$), $C_{P,v}$ is the purchase cost (\$) of vacuum pump, $\dot{W}_v$ is the maximum shaft power (kW) of vacuum pump, $C_{P,comp}$ is the purchase cost (\$) of the compressor, $C_{P,drive}$ is the purchase cost (\$) of the electric motor, and $\dot{W}_c$ is the maximum shaft power (kW) of the compressor, and $\eta$ is the efficiency of the compressor.

The bare module costs (BMC) of the adsorption columns, vacuum pumps, compressors, and compressor electric motors were calculated using eqs 29, 30, 31, and 32, respectively:

$$C_{BM,col} = (2.25 + 1.82 F_{m,col} F_{P,col}) C_{P,col} \times \frac{CEPCI_{2024}}{CEPCI_{2001}} \tag{29}$$

$$C_{BM,v} = (1.89 + 1.35 F_{m,v} F_{P,v}) C_{P,v} \times \frac{CEPCI_{2024}}{CEPCI_{2001}} \tag{30}$$

$$C_{BM,comp} = 2.7 C_{P,comp} \times \frac{CEPCI_{2024}}{CEPCI_{2001}} \tag{31}$$

$$C_{BM,drive} = 1.5 C_{P,drive} \times \frac{CEPCI_{2024}}{CEPCI_{2001}} \tag{32}$$

where $C_{BM,col}$ is the BMC (\$) of the adsorption column, $C_{BM,v}$ is the BMC (\$) of the vacuum pump, $C_{BM,comp}$ is the BMC (\$) of the compressor, $C_{BM,drive}$ is the BMC (\$) of the electric motor, $F_{m,i}$ is the material factor for adsorption column or vacuum pump, $F_{P,i}$ is the pressure factor for adsorption column or vacuum pump, and $CEPCI_i$ is the chemical engineering plant cost index (CEPCI) for year $i$. For adsorption columns constructed from carbon steel, the material factor ($F_{m,col}$) was set to 1 and the pressure factor ($F_{P,col}$) was calculated using eq 33:

$$F_{P,col} = max\left(\frac{\frac{(P_H + 1)2r_{in}}{2(850 - 0.6(P_H + 1))} + 0.00315}{0.0063}, 1.25\right) \tag{33}$$

where $P_H$ is the adsorption pressure (barg). For vacuum pumps constructed from carbon steel, the material ($F_{m,v}$) and pressure factors ($F_{P,v}$) were set to 1.6 and 1, respectively.

The total bare module cost (TBMC) was calculated as the sum of the BMCs of each equipment using eq 34:

$$TBMC\ (\$) = MN_{col}C_{BM,col} + MN_{v,Depres}C_{BM,v,Depres} + MN_{v,LR}C_{BM,v,LR} + M(C_{BM,comp,feed} + C_{BM,drive,feed}) + M(C_{BM,comp,HR} + C_{BM,drive,HR}) \quad (34)$$

where $C_{BM,v,k}$ is the BMC of the vacuum pump for the depressurization or light reflux step, and $C_{BM,comp,k}$ is the BMC of the compressor for the feed or heavy reflux step. The contingency and fee costs, assumed to be 15% and 3% of the TBMC, respectively, and were added to the TBMC to calculate the total module cost (TMC), as follows:

$$TMC\ (\$) = 1.18 \times TBMC \quad (35)$$

Additional costs associated with site development, auxiliary buildings, off-site facilities, and utilities were assumed to be 50% of the TBMC and were added to the TMC to calculate the grassroots cost ($C_{GR}$) or CAPEX, as follows:

$$C_{GR}\ or\ CAPEX\ (\$) = TMC + 0.50 \times TBMC \quad (36)$$

The CAPEX was annualized using the equivalent annual cost (EAC)[58] approach, and the annualized CAPEX (i.e., EAC) was calculated using eq 37:

$$EAC\ (\$/yr) = \frac{C_{GR} \times d}{1 - (1 + d)^{-t}} \quad (37)$$

where $d$ is the discount rate, and $t$ is the economic lifetime of the project. Economic parameters used in the estimation of EAC are provided in Table 2.

**Table 2**. **Economic parameters used in techno-economic analysis.**

| Parameter | Unit | Value | Reference |
|---|---|---|---|
| Discount rate, $d$ | - | 0.08 | [59] |
| Economic lifetime, $t$ | yr | 25 | [59] |
| Electricity unit cost, $C_{elec}$ | \$/kWh | 0.07 | [57] |
| Adsorbent cost, $C_{ads}$ | \$/kg | 1.5 | [59] |
| Chemical engineering plant cost index, CEPCI | | | |
| 2024 | - | 798.8 | - |
| 2001 | - | 397 | - |

### 2.6.2.2 Operating cost

The operating cost of the PVSA system was estimated as the sum of electricity, adsorbent replacement, labor, supervision, maintenance, operating supplies, administrative overhead, and plant overhead costs. The electricity cost ($OC_{elec}$) was calculated using eq 38:

$$OC_{elec}(\$/yr) = E_{total}Re_{CH_4}\dot{m}_{CH_4,total}C_{elec} \quad (38)$$

where $E_{total}$ is the total energy consumption (kWh/tonne $CH_4$), $\dot{m}_{CH_4,total}$ is the total mass flow rate (tonne CH4/yr) of CH4 in the natural gas, and $C_{elec}$ is the unit cost of electricity ($/kWh). The total energy consumption ($E_{total}$) was calculated using eq 39:

$$E_{total}\left(\frac{kWh}{tonne\ CH_4}\right) = \frac{E_{Pres} + E_{Ads} + E_{HR} + E_{Depres} + E_{LR}}{\left(n_{CH_4}^{Out\ from\ Ads}(1-\alpha_{LR}) + n_{CH_4}^{Out\ from\ HR}\right)M_{CH_4}} \quad (39)$$

where $M_{CH_4}$ is the molecular weight of CH4 (tonne/mol), and $E_i$ is the energy consumption (kWh) for each step in the PVSA cycle and was calculated as follows:

$$E_{Pres} = \begin{cases} \varepsilon\pi {r_{in}}^2\left(\frac{\gamma}{\gamma-1}\right)v_0P_0\int_0^{t_{Pres}}\left(\frac{\overline{v}\overline{P}}{\eta}\right)\Big|_{Z=0}\left[\left(\frac{P_0\overline{P}\big|_{Z=0}}{P_F}\right)^{\frac{\gamma-1}{\gamma}}-1\right]dt, if\ P_0\overline{P}\big|_{Z=0} > P_F \\ 0,\ \ if\ P_0\overline{P}\big|_{Z=0} < P_F \end{cases} \quad (40)$$

$$E_{Ads} = \varepsilon\pi {r_{in}}^2\left(\frac{\gamma}{\gamma-1}\right)v_0P_0\int_0^{t_{Ads}}\left(\frac{\overline{v}\overline{P}}{\eta}\right)\Big|_{Z=0}\left[\left(\frac{P_0\overline{P}\big|_{Z=0}}{P_F}\right)^{\frac{\gamma-1}{\gamma}}-1\right]dt \quad (41)$$

$$E_{HR} = \varepsilon\pi {r_{in}}^2\left(\frac{\gamma}{\gamma-1}\right)v_0P_0\int_0^{t_{HR}}\left(\frac{\overline{v}\overline{P}}{\eta}\right)\Big|_{Z=0}\left[\left(\frac{P_0\overline{P}\big|_{Z=0}}{P_F}\right)^{\frac{\gamma-1}{\gamma}}-1\right]dt \quad (42)$$

$$E_{Depres} = \begin{cases} \varepsilon\pi {r_{in}}^2\left(\frac{\gamma}{\gamma-1}\right)v_0P_0\int_0^{t_{Depres}}\left(\frac{\overline{v}\overline{P}}{\eta}\right)\Big|_{Z=0}\left[\left(\frac{P_{atm}}{P_0\overline{P}\big|_{Z=0}}\right)^{\frac{\gamma-1}{\gamma}}-1\right]dt, if\ P_0\overline{P}\big|_{Z=0} < P_{atm} \\ 0,\ \ if\ P_0\overline{P}\big|_{Z=0} > P_{atm} \end{cases} \quad (43)$$

$$E_{LR} = \varepsilon\pi {r_{in}}^2\left(\frac{\gamma}{\gamma-1}\right)v_0P_0\int_0^{t_{LR}}\left(\frac{\overline{v}\overline{P}}{\eta}\right)\Big|_{Z=0}\left[\left(\frac{P_{atm}}{P_0\overline{P}\big|_{Z=0}}\right)^{\frac{\gamma-1}{\gamma}}-1\right]dt \quad (44)$$

where $\varepsilon$ is the column void fraction, and $r_{in}$ is the column radius (m), $\gamma$ is the adiabatic constant, $\eta$ is the efficiency of the compressor/vacuum pump, and $P_F$ is the absolute pressure (Pa) of the feed gas. The adsorbent was assumed to be completely replaced every 1.5 years to account for capacity loss and related degradation from continuous cyclic operation. The cost of adsorbent replacement ($OC_{ads}$) was calculated using eq 45:

$$OC_{ads}(\$/yr) = MN_{col}V(1-\varepsilon)\rho_sC_{ads}/1.5 \quad (45)$$

where $\rho_s$ is the density of adsorbent (kg/m$^3$) and $C_{ads}$ is the adsorbent unit cost ($/kg). A total of 10 operators, including technicians, were assumed to be required for the PVSA system, with a labor rate of $34.50 per hour. The labor cost ($OC_{labor}$) was calculated using eq 46:

$$OC_{labor}(\$/yr) = Labor\ rate \times 10 \times 365 \times 24 \quad (46)$$

The supervisory cost ($OC_{supervisory}$) was calculated as 25% of the labor cost, the maintenance cost ($OC_{maintenance}$) as 10% of the TMC, and the cost of operating supplies ($OC_{supplies}$) as 20% of the maintenance cost. The administrative ($OC_{admin}$) and plant overhead costs ($OC_{plant}$) were calculated as 15% and 70% of the sum of labor, supervisory, and maintenance costs, respectively. The operating cost (OPEX) was calculated using eq 47:

$$OPEX\ (\$/yr) = OC_{elec} + OC_{ads} + OC_{labor} + OC_{supervisory} + OC_{maintenance} + OC_{supplies} + OC_{admin} + OC_{plant} \tag{47}$$

Economic parameters used in the estimation of operating costs are provided in Table 2.

**2.6.2.3 $CH_4$ production cost**

The total annual cost (TAC) of $CH_4$ production was calculated as the sum of the EAC and the OPEX using eq 48:

$$TAC\ (\$/yr) = EAC + OPEX \tag{48}$$

The $CH_4$ production cost was calculated using eq 49:

$$C_{CH_4}^{Prod}\ (\$/tonne\ CH_4) = \frac{TAC}{Re_{CH_4}\dot{m}_{CH_4,total}} \tag{49}$$

## 3. RESULTS AND DISCUSSION

### 3.1 Case selection: Energy distribution-based screening

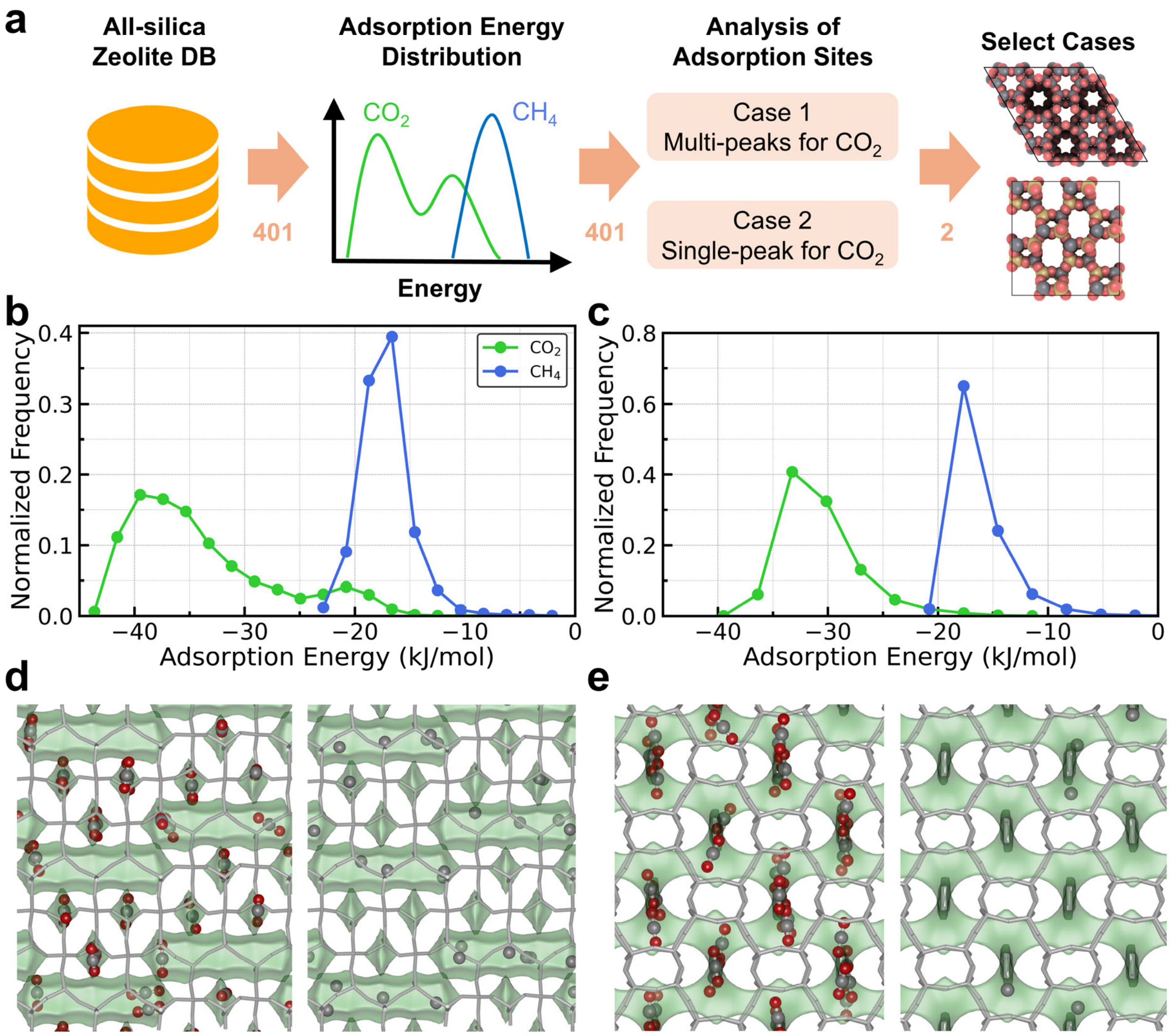

**Figure 1.** (a) Schematic representation of workflow for energy distribution-based screening. Adsorption energy distribution of a single $CO_2$ or $CH_4$ molecule inside (b) AFG-1 and (c) GIS-1 at 298 K. Molecular simulation snapshots for adsorbed $CO_2$ (left) and $CH_4$ (right) molecules in (d) AFG-1 and (e) GIS-1. Snapshots were obtained from GCMC simulations at 10 bar and 298 K under mixture condition (10/90% $CO_2/CH_4$). For clarity, the zeolite frameworks were represented by gray lines. Gray atoms indicate carbon, and red atoms indicate oxygen.

To evaluate the applicability of the MPD-based approach for process optimization, we first selected two zeolites from the all-silica zeolite database. One was chosen as a case where IAST is expected to accurately predict mixture adsorption behavior, and the other as a case where IAST is likely to fail. IAST assumes that all adsorbates have equal access to the adsorption surface; when this assumption holds, it provides accurate predictions, but deviations often cause errors[18, 22]. To identify such cases, we analyzed adsorption energy distributions of 401 zeolites in the database (Figure 1a). We hypothesize that the performance of IAST is strongly influenced by the characteristics of these energy distributions. When both gases ($CO_2$ and $CH_4$) exhibit a single distinct peak, suggesting uniform adsorption sites with equal accessibility, IAST tends to provide accurate predictions. In contrast, when the energy distribution, particularly for $CO_2$, shows multiple peaks, the presence of energetically preferred adsorption sites violates the assumptions of IAST and leads to poor predictive performance. Based on this rationale, we classified the zeolites into two categories according to the number of peaks in their $CO_2$ energy distributions. From this classification, GIS-1 was selected as a representative zeolite where IAST performs well, and AFG-1 as one where it does not. AFG-1 shows two distinct peaks for $CO_2$, suggesting the presence of two energetically distinct adsorption sites, whereas $CH_4$ exhibits only a single peak (Figure 1b). In contrast, GIS-1 shows a single adsorption site for both $CO_2$ and $CH_4$ (Figure 1c). As further shown in Figure 1d, AFG-1 clearly exhibited the presence of two $CO_2$ adsorption sites. $CO_2$ preferentially adsorbs in small pocket sites that are energetically favorable, while $CH_4$ is largely excluded from these sites, resulting in a violation of the IAST assumption of uniform site accessibility. GIS-1, on the other hand, features only one type of adsorption site (Figure 1e), equally accessible to both gases, satisfying IAST assumptions.

It is important to note that energy histogram–based classification should be interpreted with caution. This approach was used to rapidly identify two contrasting cases, where IAST is likely to perform well and where it is not, based on a physically motivated criterion. As such, we did not perform further IAST accuracy validation across all classified cases. In fact, we later found that this energy-distribution-based classification does not always perfectly correlate with IAST accuracy. Nonetheless, we believe this approach provides a reasonable and efficient means to select illustrative examples that demonstrated the advantages of MPD-based prediction method.

### 3.2 Comparison of $CO_2$/$CH_4$ mixture adsorption isotherms

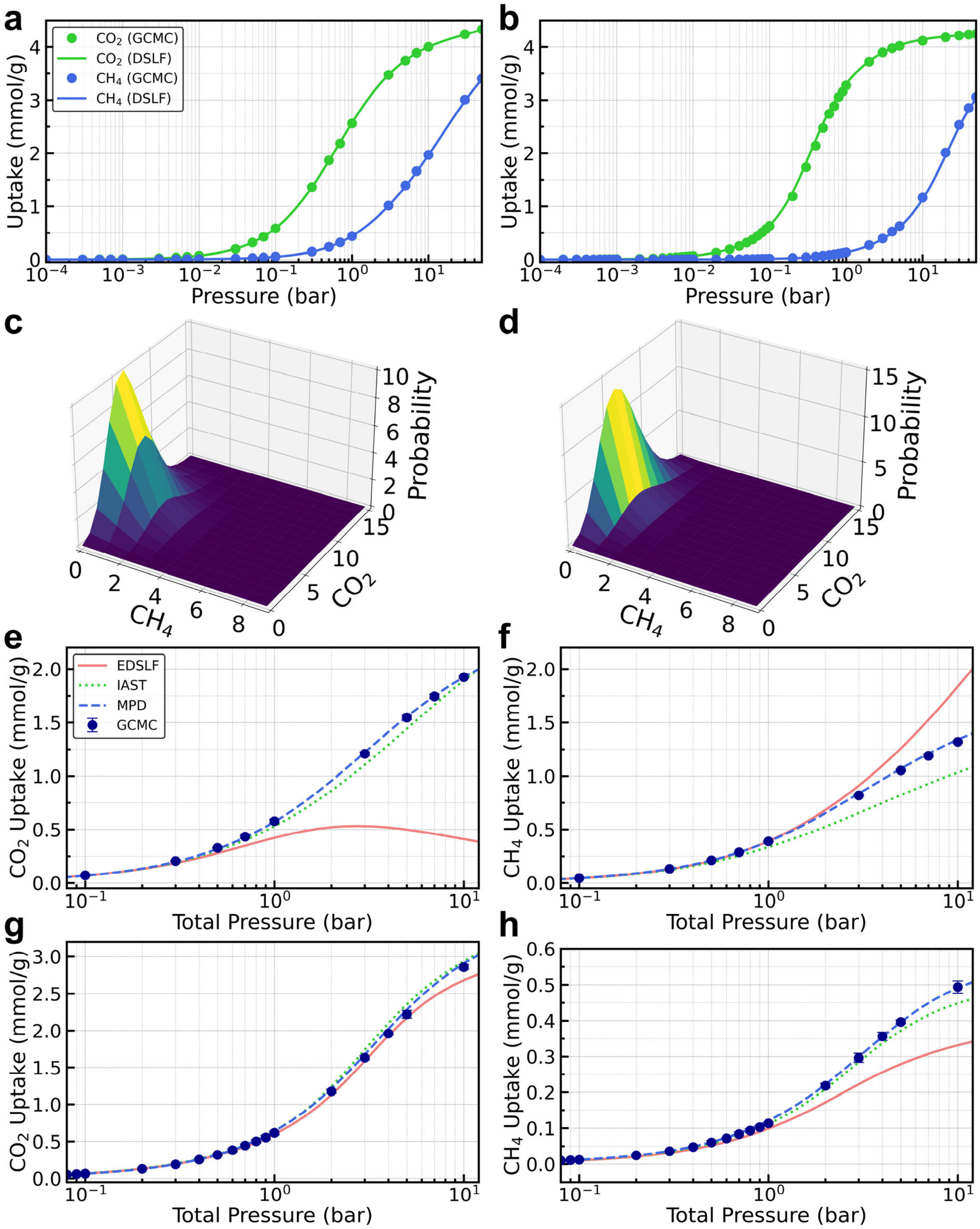

**Figure 2.** Single component adsorption isotherms and fitted models of (a) AFG-1 and (b) GIS-1 for $CO_2$ and $CH_4$ at 298 K. GCMC indicates the data obtained from GCMC simulations, and DSLF indicates the data predicted by the DSLF model. 2D MPD (%) for (c) AFG-1 and (d) GIS-1 determined by the 2D NVT+W method at reference condition (0.2 bar, 300 K, and 1:1

molar ratio). Purple and yellow colors represent low and high probabilities, respectively. Mixture adsorption isotherms of $CO_2$ (left), and $CH_4$ (right) for (e-f) AFG-1 and (g-h) GIS-1 at 298 K with 10/90% $CO_2/CH_4$ mixture. EDSLF indicates the data predicted by EDSLF model, IAST indicates the data predicted by IAST method, MPD indicates the data predicted by MPD-based approach, and GCMC indicates the data obtained from GCMC simulations under mixture conditions.

The accuracy of the MPD-based approach in predicting mixture adsorption equilibrium was evaluated against GCMC simulations, alongside predictions from two conventional approaches, EDSLF and IAST. To enable predictions using both the EDSLF and IAST methods, single-component adsorption isotherms of $CO_2$ and $CH_4$ were obtained from GCMC simulations at 298 K and subsequently fitted to the DSLF model (Figure 2a and 2b). The fitted DSLF parameters and corresponding $R^2$ values for each zeolite are listed in Table S4, and the isosteric heat of adsorption for the gases in each zeolite are presented in Table S5. Additionally, the 2D MPDs for the selected zeolites were generated using the 2D-NVT+W method (Figure 2c and 2d) at reference condition of 0.2 bar, 300 K, and 1:1 molar ratio. With the fitted DSLF parameters and 2D MPDs, binary adsorption isotherms were predicted across different temperatures and compositions using the EDSLF model, IAST method, and the MPD-based approach (Figure 2e-h, and Figures S1-S4).

At 298 K and a 10/90% $CO_2/CH_4$ feed composition, IAST underestimated the uptakes of the more weakly adsorbed $CH_4$ in AFG-1 above 1 bar (Figure 2f). This deviation arises from IAST's assumption that $CH_4$ molecules compete with all $CO_2$ molecules across the adsorption sites. As noted above, $CO_2$ preferentially occupies small pocket sites that are largely noncompetitive for $CH_4$ (Figure 1d). As a result, the extent of competition faced by $CH_4$ at those sites in AFG-1 is minimal, leading IAST to underpredict its uptake. This discrepancy became even more prominent under conditions favoring greater $CH_4$ adsorption, as IAST assumes stronger competition between components in such cases. At the same feed composition and 273 K, IAST began to underestimate $CH_4$ uptake from pressures as low as 0.3 bar, and the deviation grew larger (Figure S1b). Furthermore, IAST overestimated $CO_2$ uptake at pressures above 2 bar. In contrast, at 323 K, the deviation between IAST and GCMC predictions was significantly reduced (Figure S2a-b). A similar trend was observed with varying feed compositions. As the $CO_2$ fraction increased, the amount of $CH_4$ adsorbed in the mixture decreased, thereby reducing the extent of competition and improving IAST accuracy. At a 50/50% $CO_2/CH_4$ feed composition, the deviation became smaller (Figure S3a-b), and at 90/10%, IAST accurately predicted the mixture adsorption isotherms (Figure S4a-b). By contrast, GIS-1, which satisfies the assumptions of IAST, exhibited excellent agreement between IAST and GCMC predictions across all pressures, temperatures, and feed compositions for the $CO_2/CH_4$ mixture, with only minor deviations observed (Figure 2g-h, Figure S1c-d, Figure S2c-d, Figure S3c-d, and Figure S4-c-d). These observations underscore a fundamental limitation of IAST: its predictive accuracy can vary substantially depending on whether the adsorbent–mixture combination satisfies the underlying assumptions of the model. Since such validity often requires molecular-level insights from simulations, the use of IAST in material evaluation, particularly in high-throughput screening (HTS) studies may lead to misleading or erroneous results.

The EDSLF model exhibited similar trends to IAST but with consistently larger deviations, particularly showing severe inaccuracies for AFG-1. At a 10/90% $CO_2/CH_4$ feed composition, the EDSLF model failed to accurately capture the mixture adsorption behavior of $CO_2/CH_4$ in AFG-1, erroneously estimating $CH_4$ uptake to be higher than that of $CO_2$ (Figure 2e-f, Figure S1a-b, and Figure S2a-b). Although this large discrepancy was alleviated as the $CO_2$ fraction in the feed increased, the deviations persisted. For instance, at a 50/50% $CO_2/CH_4$ feed composition, the EDSLF model provided reasonable predictions up to 1 bar, but at higher pressures, it significantly underestimated $CO_2$ uptake while overestimating $CH_4$ uptake, resulting in much larger errors than those observed with IAST (Figure S3a-b). Even at a 90/10% $CO_2/CH_4$ feed composition, where the $CO_2$ uptake was well captured, the model continued to overestimate $CH_4$ uptake at elevated pressures (Figure S4a-b). In contrast, for GIS-1, the EDSLF model decently produced the mixture adsorption isotherms under most conditions, though relatively minor deviations were observed at the 10/90% feed composition (Figure 2g-h, Figure S1c-d, Figure S2c-d, Figure S3c-d, and Figure S4-c-d). Notably, despite the use of perfectly fitted DSLF parameters (Table S4), the accuracy of the EDSLF model varied depending on the system. In practice, multiple parameter combinations can yield excellent single-component isotherm fits within the DSLF framework, yet only a subset of these leads to accurate mixture adsorption predictions. Farmahini et al.[13] and Wilkins and Rajendran[60] proposed fitting protocols to identify such parameter sets for $CO_2/N_2$ mixture. However, these procedures have not been validated for other mixtures such as $CO_2/CH_4$. In particular, $CO_2$ and $CH_4$ can exhibit different saturation capacities; accordingly, protocols that assume equal saturation loadings for both components are not generally transferable. Developing a fitting strategy that is broadly transferable across mixtures and adsorbents is therefore nontrivial, highlighting a limitation of the EDSLF model.

Distinct from the two conventional approaches, the MPD-based approach consistently provided accurate predictions of the $CO_2/CH_4$ mixture adsorption isotherms across all pressures, temperatures, and feed compositions for both AFG-1 and GIS-1 (Figure 2e-h, and Figures S1-S4). Notably, the pressure range considered (0.1–10 bar) spans from the minimum desorption pressure to the maximum adsorption pressure defined in our process optimization, while the temperature range (273–323 K) spans the lowest and highest values observed in our previous process optimization study[15]. These results indicate that the MPD-based approach offers accurate predictions over the full range of pressures, temperatures, and compositions encountered within the adsorption column in this study. While the method was previously validated for ethane/ethylene mixtures[30], the present study further demonstrates its successful application to the $CO_2/CH_4$ system, further confirming its general applicability across different adsorbent–mixture combinations.

### 3.3 Comparison of process-level performance

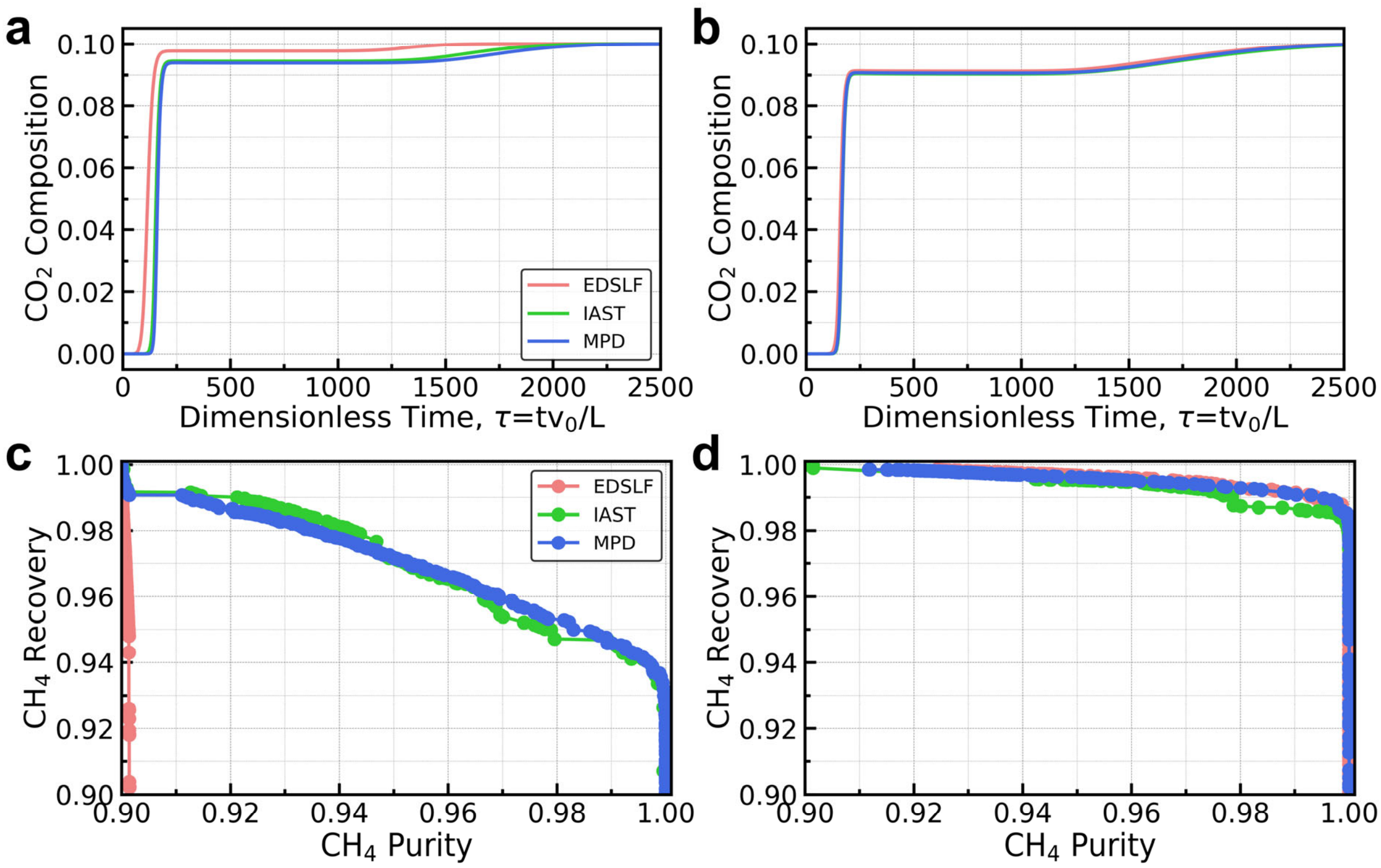

**Figure 3.** $CO_2$ composition breakthrough curves on (a) AFG-1 and (b) GIS-1 at the exit of the column with 10/90% $CO_2/CH_4$ feed mixture at 2 bar and 298 K. $CH_4$ purity/recovery Pareto fronts for (c) AFG-1 and (d) GIS-1 obtained from the EDSLF-based, IAST-based, and MPD-based process cycle optimizations for 10/90% $CO_2/CH_4$ mixture. EDSLF, IAST, and MPD denote the results obtained from EDSLF-, IAST-, and MPD-based process models, respectively.

We integrated three mixture equilibrium prediction methods into process models to evaluate the applicability of the MPD-based approach at the process level. Before full process cycle optimization, breakthrough profiles obtained from the EDSLF-, IAST-, and MPD-based process models were compared (Figure 3a-b and Figure S5). In a breakthrough simulation, only a single adsorption step is considered within the cycle, meaning that the system of ODEs needs to be solved just once per run from a modeling perspective. This approach provides an efficient platform for preliminary sensitivity analyses or timing assessments prior to full-cycle simulations or optimizations. We considered the results of the MPD-based process model as the ground truth. For AFG-1, the IAST method underestimated $CO_2$ uptake in the binary mixture at 2 bar, 298 K, and a 10/90% feed composition (Figure 2e). This underestimation led to an earlier first transition in the breakthrough curve, shifting from approximately 165 s in the MPD-based model to 150 s in the IAST-based model (Figure 3a, Figure S5a, and Figure S5c). The EDSLF model significantly underpredicted the $CO_2$ uptake under same conditions (Figure 2e), yielding a first transition at approximately 80 s, considerably earlier than that of the IAST-based model (Figure 3a, Figure S5a, and Figure S5c). For GIS-1, both IAST and MPD-based methods accurately predicted the mixture adsorption isotherms (Figure 2g), leading to identical breakthrough profiles (Figure 3b, Figure S5b, and Figure S5d). By contrast, the EDSLF model exhibited minor deviations in isotherm predictions (Figure 2g), which shifted the first transition slightly earlier, from approximately 175 s in the MPD-based model to about 150 s in the EDSLF-based model (Figure 3b, Figure S5b, and Figure S5d). Importantly, MPD-based breakthrough simulation was 3–7 times slower than that using the EDSLF model but 9–18 times faster than that using IAST (Figure S6 and see Supporting Information 1.1 for details). Although EDSLF offers faster computations, it predicts breakthrough curves with notable inaccuracies. In this context, the MPD approach presents a significant advantage by enabling breakthrough simulations at speeds substantially faster than the widely used IAST method while maintaining higher accuracy.

We extended the analysis to process optimization to assess the applicability of the MPD-based optimization approach. Consistent with the breakthrough results, the $CH_4$ purity/recovery Pareto front for AFG-1 obtained from EDSLF-based optimization significantly deviated from those obtained from IAST and MPD-based optimizations (Figure 3c). As previously shown, the EDSLF model failed to capture the correct separation behavior, predicting a higher $CH_4$ uptake than $CO_2$ at a 10/90% feed composition (Figure 2e-f and Figures S1-S2). This leads to no effective $CH_4$ separation. As with the breakthrough simulations, the MPD-based optimization results were regarded as the ground truth. Similar to findings from the breakthrough simulations, the IAST-based optimization produced a $CH_4$ purity/recovery Pareto front comparable to that of the MPD model, with only slight deviations observed. The optimal variable distributions from the IAST- and MPD-based process cycle optimizations, which produced similar $CH_4$ purity/recovery Pareto fronts, largely overlapped. Distinctly, as shown in Figure S7, the distributions of optimal decision variables obtained from the EDSLF-based process cycle optimization, which failed to achieve effective $CH_4$ separation, differed markedly, particularly for the light reflux ratio and desorption pressure. For GIS-1, in line with previous results for breakthrough profiles, the optimizations based on the three methods produced nearly identical $CH_4$ purity/recovery Pareto fronts. (Figure 3d). Aside from minor differences in the distribution of optimal adsorption pressures, the overall distributions of optimized decision variables obtained from the optimizations were comparable across all three methods (Figure S8).

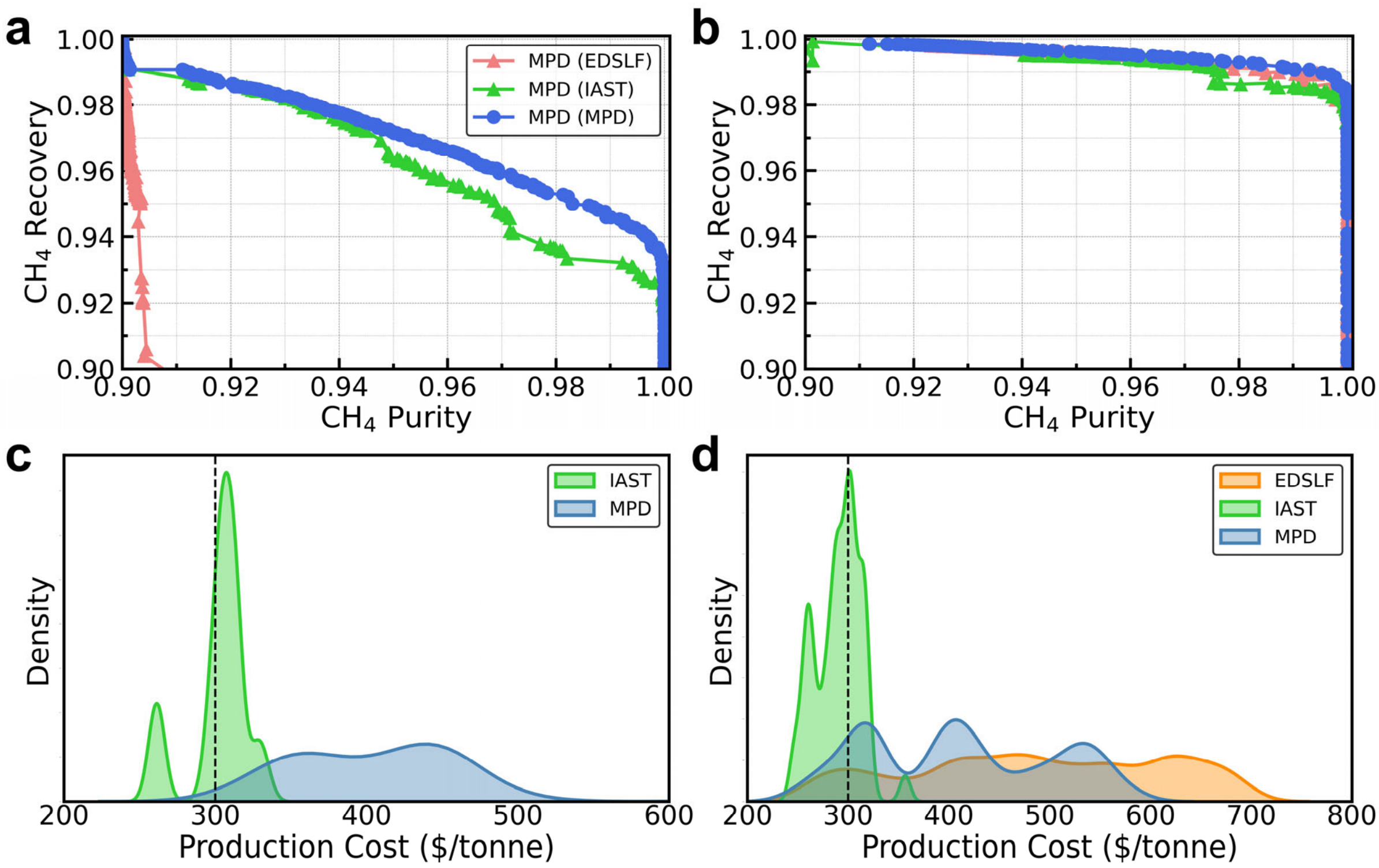

**Figure 4.** $CH_4$ purity/recovery Pareto fronts for (a) AFG-1 and (b) GIS-1 recalculated using the MPD-based process model with operating conditions from the EDSLF-, IAST-, and MPD-based optimizations. MPD(EDSLF), MPD(IAST), and MPD(MPD) indicate the Pareto fronts recalculated using the MPD-based process model with operating conditions from the EDSLF-, IAST-, and MPD-based optimizations, respectively. Distribution of $CH_4$ production cost for (c) AFG-1 and (d) GIS-1 at pipeline-quality purity (≥98%) calculated from each variable sets. The black dashed lines indicate the minimum natural gas price in Europe[61].

To assess whether the operating conditions from each optimization were truly optimal, we re-evaluated the optimal decision variable sets from the EDSLF- and IAST-based optimizations. This was done using the MPD-based process model to reconstruct the $CH_4$ purity/recovery Pareto fronts. As shown in Figure 4a, the variable sets from the EDSLF-based optimization, which poorly captured the mixture adsorption behavior, failed to reproduce the Pareto front obtained from the MPD model. Although the Pareto front predicted by the IAST-based optimization closely matched that of the MPD-based optimization, the actual front differed noticeably when re-evaluated with the MPD-based model (Figure 4a). This discrepancy arises because the IAST-based optimization identified operating conditions that were optimal for the IAST model, not for the real adsorption behavior represented by the MPD model. For AFG-1, the IAST model failed to accurately predict the $CH_4$ mixture isotherms, causing its selected variables to deviate from the true optimum. For GIS-1, all models predicted the mixture adsorption isotherms with reasonable accuracy. Consequently, Pareto fronts recalculated with MPD using variables from the EDSLF- and IAST-based optimizations still matched the MPD-based front closely (Figure 4b).

To further assess the economic impacts of the optimal operating conditions, $CH_4$ production cost calculations were performed for each variable set which satisfies pipeline-quality $CH_4$ purity (≥98%) (see Methods 2.6 for details). For AFG-1, the production cost distribution for IAST-based variables is noticeably left-shifted relative to that for MPD-based variables (Figure 4c). As a result, the IAST-based variables yielded a minimum cost of \$260.6 per tonne $CH_4$, compared with \$308.7 per tonne $CH_4$ for the MPD-based variables. In contrast, the production cost distributions for GIS-1 (Figure 4d) overlapped substantially across all variable sets, with minimum costs of \$261.9, \$246.4, and \$246.4 per tonne $CH_4$ for EDSLF, IAST, and MPD-based variables, respectively. The small (~6%) deviation observed for EDSLF-based variable stemmed from its modest adsorption prediction error. While full cost-optimization was beyond the scope of this study, these results demonstrated that even when a model produces a seemingly accurate Pareto front, errors in mixture adsorption predictions can lead to suboptimal operating conditions. Such errors can misrepresent both process performance and economic potential. Most large-scale computational screening studies have incorporated simplified models, such as the dual-site Langmuir model, directly into process modeling due to their computational efficiency[62-66]. Our findings suggest that, without verified accuracy in mixture adsorption predictions, such approaches can yield misleading conclusions when assessing material performance in practical separation processes. These observations underscore the critical need for accurate mixture adsorption modeling to ensure reliable process evaluation and material screening.

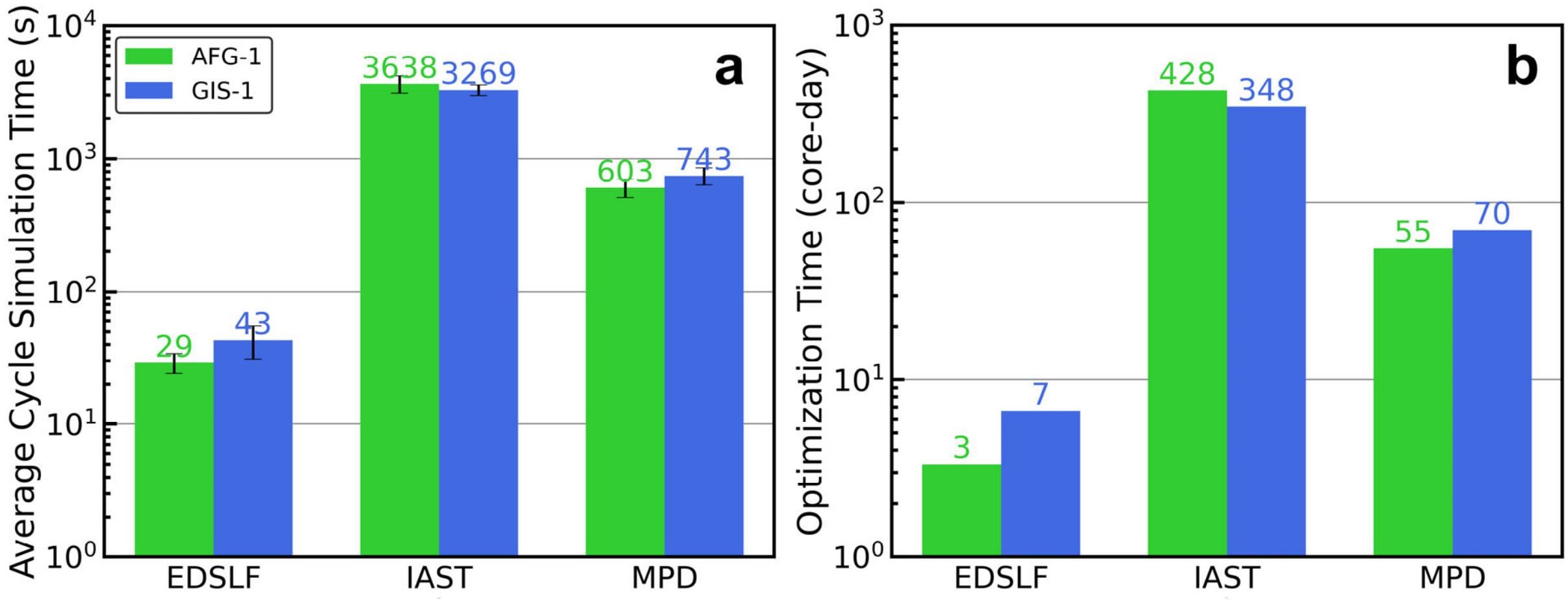

**Figure 5.** Comparison of (a) average cycle simulation time and (b) total optimization time for EDSLF-based, IAST-based, and MPD-based process cycle optimizations. Core-days were calculated from the number of CPU cores multiplied by the elapsed wall time.

Figure 5 summarizes the computational load associated with the incorporation of different mo dels in process optimization. Similar to those observed in breakthrough simulations, MPD-based optimization was 10–18 times slower than EDSLF-based optimization but 5–8 times faster than IAST-based optimization. Unlike breakthrough simulations, which require solving a complex ODE system only once, process simulations require repeatedly solving the ODE system until cyclic steady state is reached. In these cases, the MPD method was substantially slower than EDSLF, yet remained significantly faster than IAST. While MPD-based optimizations were completed within four days (55 core-days for AFG-1 and 70 core-days for GIS-1), the IAST-based optimizations required two to three weeks (428 core-days for AFG-1 and 348 core-days for GIS-1) to complete. These results suggest that the MPD-based approach, aside from its superior accuracy, also represents a practical and efficient route to the process optimization of binary adsorption systems.

To further support the generality of our conclusions, we selected two additional zeolites, one from each category. The selected zeolites were LIO-1 and APC-0, where the former exhibits two peaks in the $CO_2$ adsorption energy distribution, whereas APC-0 shows a single peak for both $CO_2$ and $CH_4$ (Figure S9). As a result, IAST failed to accurately predict the mixture adsorption isotherms of LIO-1 but performed well for APC-0 (Figure S10). Likewise, the EDSLF model failed to accurately capture the $CO_2$/$CH_4$ mixture adsorption behavior for LIO-1, while yielding accurate predictions for APC-0 (Figure S10). In contrast, the MPD-based approach accurately reproduced the mixture adsorption isotherms for both zeolites (Figure S10). Details of the fitted models, 2D MPDs, fitted DSLF parameters, corresponding $R^2$ values, and isosteric heat of adsorption for the gases in each zeolite are provided in Figure S11 and Tables S6–S7. We subsequently carried out breakthrough simulations, as in the earlier analysis. For LIO-1, the first transition in the breakthrough curve shifted depending on the degree of underestimation in the predicted isotherms (Figure S12), consistent with the behavior observed for AFG-1. For APC-0, all three methods produced essentially identical breakthrough curves (Figure S12). Comparison of computational times showed that, for these additional zeolites, MPD-based process simulations remained approximately 4–13 times slower than those using EDSLF, yet still 6–10 times faster than those using IAST (Figure S13). Together, these results demonstrate that the superiority of the MPD-based approach, in both accuracy and computational efficiency, extends beyond the original two zeolites and is generally applicable across diverse adsorbent frameworks.

### 3.4 Extension to ternary mixture

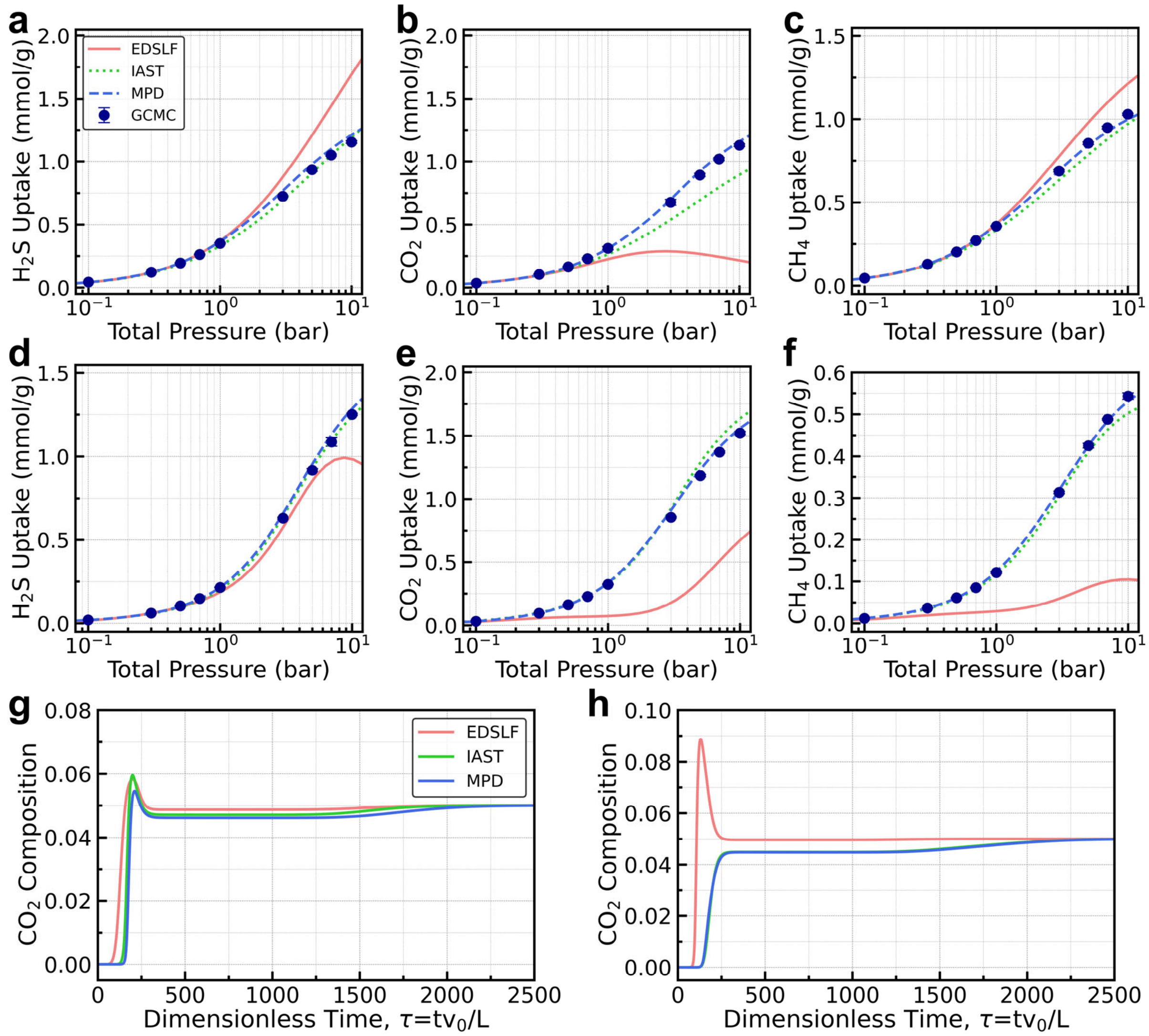

**Figure 6.** Mixture adsorption isotherms of $H_2S$ (left), $CO_2$ (middle), and $CH_4$ (right) for (a-c) AFG-1 and (d-f) GIS-1 at 298 K with 5/5/90 % $H_2S/CO_2/CH_4$ mixture. EDSLF indicates the data predicted by EDSLF model, IAST indicates the data predicted by IAST method, MPD indicates the data predicted by MPD-based approach, and GCMC indicates the data obtained from GCMC simulations under mixture conditions. $CO_2$ composition breakthrough curves on (g) AFG-1 and (h) GIS-1 at the exit of the column with 5/5/90% $H_2S/CO_2/CH_4$ feed mixture at 2 bar and 298 K. EDSLF, IAST, and MPD denote the results obtained from EDSLF-, IAST-, and MPD-based breakthrough simulations, respectively.

To assess the performance of the MPD-based approach in multicomponent adsorption systems, we extended our analysis from binary ($CO_2/CH_4$) to ternary mixtures ($H_2S/CO_2/CH_4$). As in the binary case, we first generated ternary mixture adsorption isotherms at three different temperatures using three different models: EDSLF, IAST, and the MPD-based approach (i.e., with the extended, 3D NVT+W approach, developed herein with details shown in Methods 2.3). In AFG-1, the EDSLF model again failed to capture the correct adsorption behavior of ternary mixture, and IAST also showed poor agreement with the reference data for ternary systems (Figure 6a-c, Figure S14a-c, and Figure S15a-c). In contrast, the 3D MPD-based approach accurately predicted ternary mixture adsorption isotherms under all tested conditions. Interestingly, results for GIS-1 differed from those observed in the binary case. Similar to AFG-1, the EDSLF model exhibited significant deviations in predicting ternary adsorption behavior of GIS-1 (Figure 6d-f, Figure S14d-f, and Figure S15d-f). In this case, IAST also showed minor discrepancies. Nevertheless, the 3D MPD-based approach provided highly accurate predictions across all conditions. These results confirm that the NVT+W framework is broadly applicable and accurate even for complex multicomponent adsorption systems.

As in the previous case, breakthrough simulations were performed using process models based on the three different methods to further evaluate the applicability of the MPD-based approach to the ternary system (Figure 6g-h and Figure S16). Here, the result from MPD-based process model was again considered as the ground truth. For AFG-1, the IAST method underestimated the uptake of $CO_2$ in the ternary mixture at 2 bar and 298 K (Figure 6b). This underestimation shifted the first transition in the breakthrough curve from approximately 195 s in the MPD-based model to about 170 s in the IAST-based model (Figure 6g and Figure S16a). The EDSLF model exhibited an even larger underestimation (Figure 6b), which produced a first transition near 100 s, substantially earlier than that of the IAST-based model (Figure 6g and Figure S16a). Overall, the predictive trends for AFG-1 were consistent with those observed for the binary system. For GIS-1, the IAST method accurately predicted the $CO_2$ uptake in the ternary mixture at 2 bar and 298 K (Figure 6e), yielding a first transition (~175 s) nearly identical to that from the MPD-based model (Figure 6h and Figure S16b). This was also consistent with the trends seen in the binary system. However, in contrast to the binary case, the EDSLF model severely underestimated the $CO_2$ uptake in the ternary mixture for GIS-1 (Figure 6e). This led to a much earlier first transition at approximately 100 s and markedly different breakthrough profiles (Figure 6h and Figure S16b).

The most pronounced difference from the binary case was observed in computational efficiency. For ternary mixtures, MPD-based simulations remained slower than those using EDSLF. Unlike the binary case, they were not faster than those using IAST and instead exhibited comparable speeds (Figure S17). This slowdown is attributed to the substantial increase in the number of elements in the MPD matrix when extending from 2D to 3D (Supporting Information 1.2). Our previous work demonstrated that IAST-based process optimization for ternary mixtures required more than one month to complete[15], which is clearly impractical. Given the similar computational cost expected for MPD-based process optimization in ternary systems, despite its superior accuracy, this approach may no longer represent the most viable option in such cases. To enable practical application of the MPD framework for multicomponent process design, future research should focus on reducing the dimensionality of the MPD matrix or developing more efficient reweighting algorithms tailored to high-dimensional systems.

## 4. CONCLUSIONS

In this study, we have introduced a material-to-process modeling framework that couples macrostate probability distributions (MPDs) from the flat-histogram Monte Carlo method with rigorous process simulation and optimization for adsorption separation applications. We demonstrated that binary and ternary adsorption systems can be predicted with high accuracy and at substantially reduced computational cost compared to IAST. These findings highlight both the potential of MPD-based methods for adsorption process design and optimization and the opportunities for further improvements, including dimensionality reduction and faster reweighting strategies for higher-dimensional systems.

The integration of MPD-based predictions into process modeling removes the need for repeated mixture simulations or extensive parameter fitting, enabling reliable equilibrium predictions across wide ranges of pressures, temperatures, and compositions. This capability is particularly powerful for high-throughput, multi-scale materials discovery campaigns, where thousands of candidate materials must be screened under realistic process conditions. In carbon capture, for example, large-scale screening studies have shown that material rankings can change significantly once realistic process models are applied, and that errors in mixture equilibrium predictions, such as those that can arise in EDSLF or IAST approaches, can misdirect discovery efforts.

By providing a physically rigorous and reweightable equilibrium model, the MPD framework ensures that screening results remain reliable across the full range of operating conditions, accelerating the identification of top-performing materials. This approach is equally applicable to other critical separations such as hydrogen purification, olefin/paraffin separation, and water harvesting, where coupling predictive thermodynamics with process-level metrics is essential for translating material discovery into deployable technologies. Moreover, with polarizable or reactive force fields to capture strong specific interactions, the MPD framework can be extended to strongly polar and chemisorptive systems, establishing a pathway toward greater generality of the approach.

## ASSOCIATED CONTENT

### Supporting Information

The Supporting Information is available free of charge at ### site address ###.

> Force field parameters, isotherm parameters, process models, isotherm data, distributions of optimized variables

## AUTHOR INFORMATION

### Corresponding Authors

**Li-Chiang Lin** − *William G. Lowrie Department of Chemical and Biomolecular Engineering, The Ohio State University, Columbus, Ohio 43210-1350, United States; Department of Chemical Engineering, National Taiwan University, Taipei 106319, Taiwan*; orcid.org/0000-0002-2821-9501; Email: lclin@ntu.edu.tw

**Yongchul G. Chung** − *School of Chemical Engineering, Pusan National University, Busan 46241, Republic of Korea; Graduate School of Data Science, Pusan National University, Busan, 46241, Republic of Korea;* orcid.org/0000-0002-7756-0589; Email: drygchung@gmail.com

### Authors

**Sunghyun Yoon** − *School of Chemical Engineering, Pusan National University, Busan 46241, Republic of Korea*

**Jui Tu** − *Department of Chemical Engineering, National Taiwan University, Taipei 106319, Taiwan*

### Author Contributions

**S.Y.** and **J.T.** contributed equally to this work. **S.Y.**: methodology (lead), formal analysis, investigation. data curation, writing—original draft, writing—review & editing, visualization. **J.T.**: investigation, data curation, writing-original draft. **L.-C.L.**: conceptualization, methodology, formal analysis, investigation, resources, writing—original draft, writing—review & editing, supervision, project administration, funding acquisition. **Y.G.C.**: conceptualization, methodology, formal analysis, investigation, resources, writing—original draft, writing—review & editing, supervision, project administration, funding acquisition.

### Notes

The authors declare no competing financial interest.

## ACKNOWLEDGEMENTS

S.Y and Y.G.C. acknowledge the National Research Foundation of Korea (NRF) for financial support from a grant funded by the Korea government (MSIT) (RS-2021-NR057842, RS-2023-00242528, RS-2024-00449431). The authors acknowledge the computational resources provided by KISTI (KSC-2024-CRE-0283). J.T. and L.-C.L. acknowledge financial support fr om Yushan Fellow Program by the Ministry of Education in Taiwan (MOE-110-YSFEE-0003

-002-P1) and the National Science and Technology Council (113-2223-E-002-008-MY5 and 1 13-2628-E-002-016-MY3), as well as computational resources from the National Center for H igh-performance Computing (NCHC) in Taiwan.

**Data and Software Availability Statement**

The calculated mixture isotherms, macrostate probability distributions, and process optimization results, as well as simulation codes and output files, are available at https://github.com/Chung-Research-Group/reproducible-workflows/tree/master/2025-MPD-PVSA.

## TOC Graphic

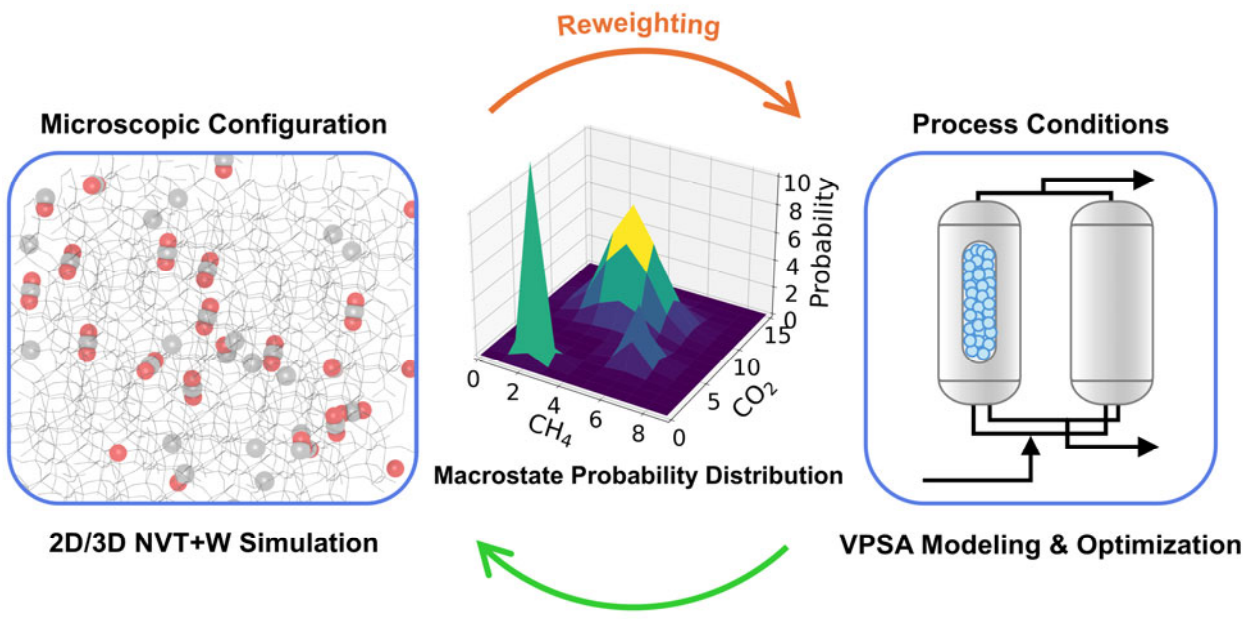
Reweighting
Microscopic Configuration
Process Conditions
Probability
$CH_4$
$CO_2$
Macrostate Probability Distribution
2D/3D NVT+W Simulation
VPSA Modeling & Optimization

## Supporting Information:

# Integrating Macrostate Probability Distributions with Swing Adsorption Modeling for Binary/Ternary Gas Separation

Sunghyun Yoon[1,5], Jui Tu[2,5], Li-Chiang Lin[2,3,*], and Yongchul G. Chung[1,4,*]

[1]*School of Chemical Engineering, Pusan National University, Busan, 46241, Republic of Korea*

[2]*Department of Chemical Engineering, National Taiwan University, Taipei 106319, Taiwan*

[3]*William G. Lowrie Department of Chemical and Biomolecular Engineering, The Ohio State University, Ohio 43210, USA*

[4]*Graduate School of Data Science, Pusan National University, Busan, 46241, Republic of Korea*

[5]*These authors contribute equally*

*Corresponding authors.

E-mail: lclin@ntu.edu.tw (L.-C. Lin), drygchung@gmail.com (Y.G. Chung)

**Table of Contents**

## 1. Computational efficiency comparison of different implementation strategies for MPD reweighting

### 1.1. Binary system

The reweighting of the MPD involves applying a mathematical expression to each element of the 2D MPD matrix, and the computational efficiency of this step is highly dependent on the implementation strategy. As this directly affects the computational cost of process optimization, we conducted breakthrough simulations to compare the computational efficiency of different reweighting implementations. Three reweighting strategies were evaluated in this work: (1) an explicit loop-based implementation, (2) a fully vectorized implementation without explicit loops, and (3) a hybrid approach combining both. For comparison, breakthrough simulations using the conventional EDSLF and IAST models were also performed under the same conditions. All MPD-based breakthrough simulations were slower than those using EDSLF model but faster than those using IAST, which requires solving a system of nonlinear equations (Figure S6). Although the MPD-based approach also employs an explicit analytic expression, similar to EDSLF, the reweighting of all elements in the 2D MPD matrix introduced a non-negligible computational cost. Among the three strategies, the vectorized implementation demonstrated the best performance, being only 3–7 times slower than EDSLF, but 9–18 times faster than IAST, while the loop-based implementation was the most time-consuming. As a result, the vectorized implementation for reweighting MPD was selected for all subsequent process optimization tasks. Note that all implementations yielded identical breakthrough profiles.

### 1.2. Ternary system

In contrast to the binary systems, the computational efficiency of the MPD-based approach changed substantially in the ternary system. Notably, none of the MPD reweighting implementations outperformed the IAST-based model in terms of computational speed (Figure S17). The loop-based reweighting strategy was more than 30 times slower than IAST-based model, while even the vectorized implementation, which was the most efficient in binary systems, was still 2 to 3 times slower. Among all MPD reweighting implementations, the hybrid approach offered the fastest performance in the ternary system, but its speed was comparable to, and not faster than, that of IAST-based model. This significant drop in computational efficiency stems from the rapid increase in the number of elements in the MPD matrix as the system dimensionality increases. For binary mixtures, the 2D MPDs contained approximately 2,500 elements for both zeolites. However, in the ternary systems, the 3D MPDs for both materials included more than 140,000 elements. Consequently, the time required for reweighting 3D MPD far exceeds that needed to solve the nonlinear equations in IAST, highlighting a fundamental limitation of the MPD-based approach. A potential solution to this problem includes development of condition-dependent reweighting method which only reweight parts of the matrix that are relevant (i.e., non-zeros) but the development of this approach is beyond the scope of this work.

## 2. Supplementary figures

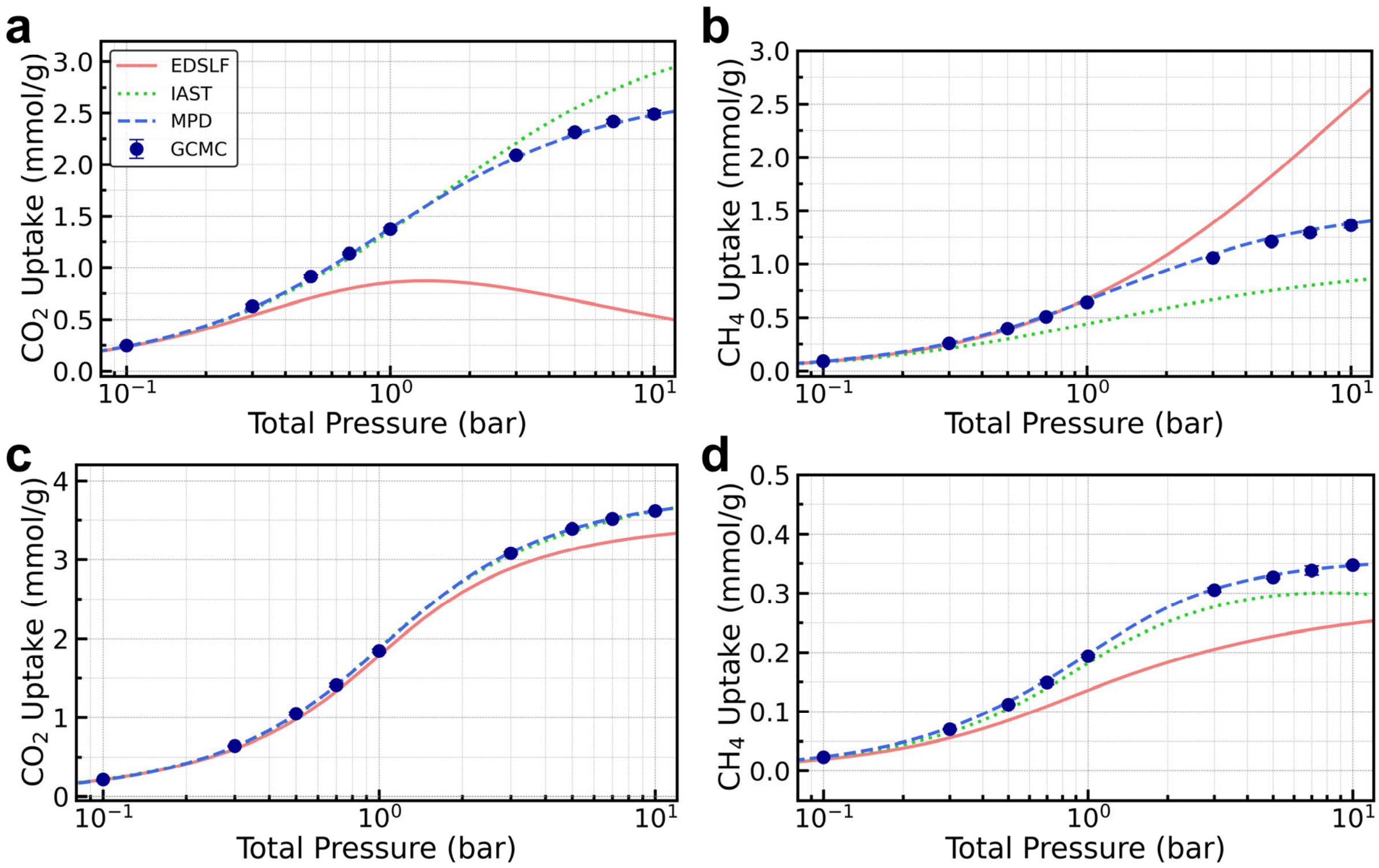

**Figure S1.** Mixture adsorption isotherms of $CO_2$ (left), and $CH_4$ (right) for (a-b) AFG-1 and (c-d) GIS-1 at 273 K with 10/90% $CO_2$/$CH_4$ mixture. EDSLF indicates the data predicted by EDSLF model, IAST indicates the data predicted by IAST method, MPD indicates the data predicted by MPD-based approach, and GCMC indicates the data obtained from GCMC simulations under mixture conditions.

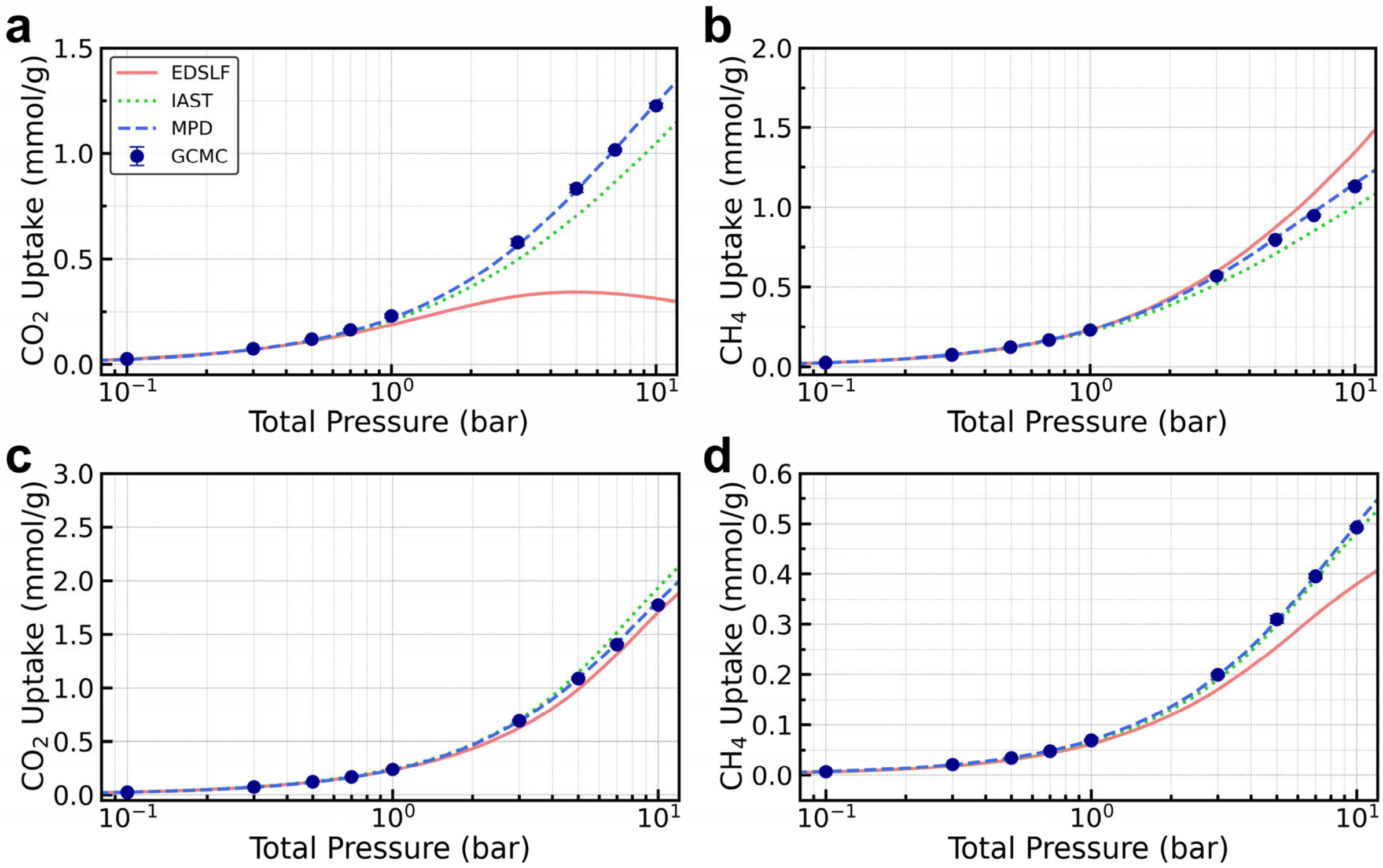

**Figure S2.** Mixture adsorption isotherms of $CO_2$ (left), and $CH_4$ (right) for (a-b) AFG-1 and (c-d) GIS-1 at 323 K with 10/90% $CO_2/CH_4$ mixture. EDSLF indicates the data predicted by EDSLF model, IAST indicates the data predicted by IAST method, MPD indicates the data predicted by MPD-based approach, and GCMC indicates the data obtained from GCMC simulations under mixture conditions.

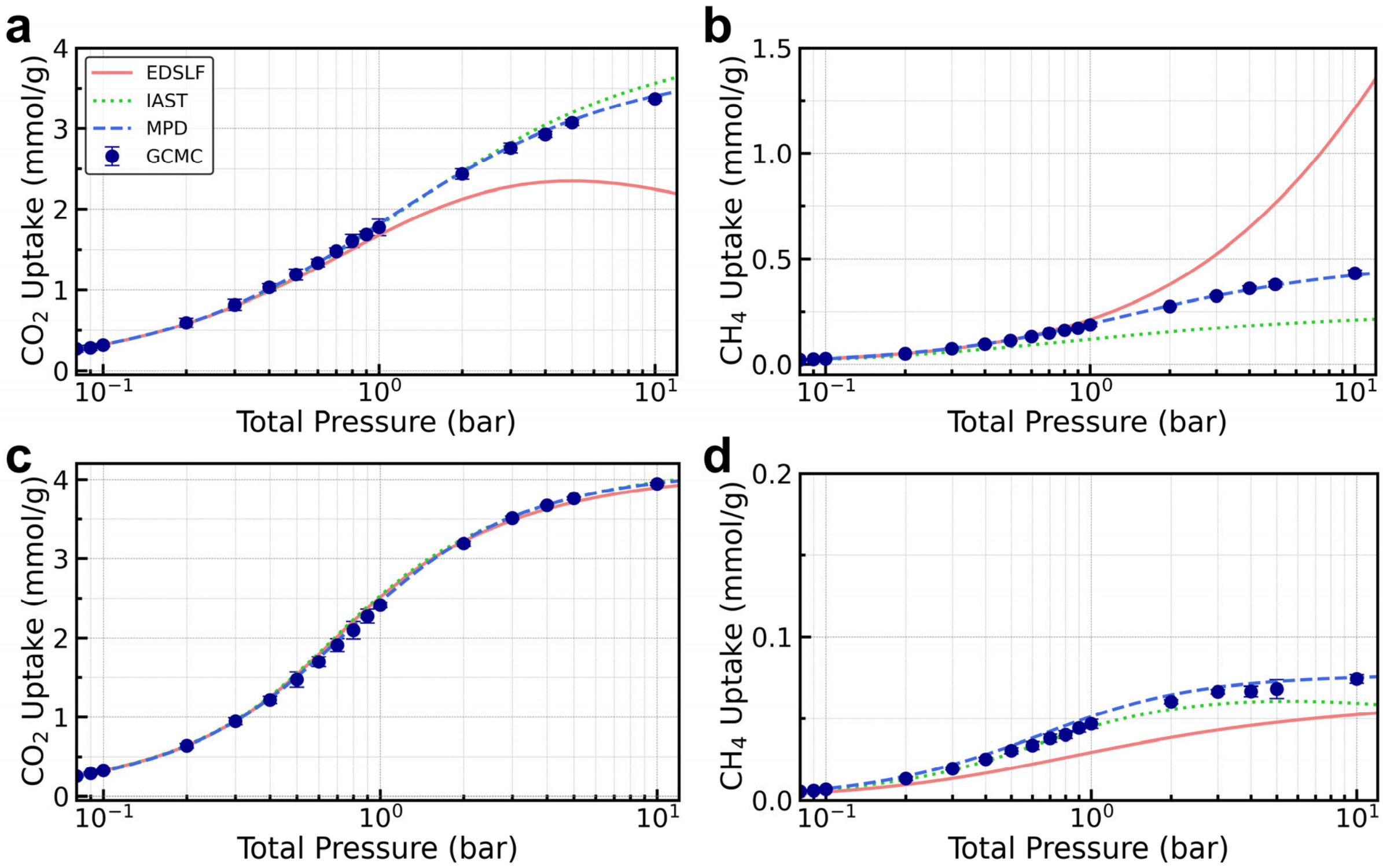

**Figure S3.** Mixture adsorption isotherms of $CO_2$ (left), and $CH_4$ (right) for (a-b) AFG-1 and (c-d) GIS-1 at 298 K with 50/50% $CO_2/CH_4$ mixture. EDSLF indicates the data predicted by EDSLF model, IAST indicates the data predicted by IAST method, MPD indicates the data predicted by MPD-based approach, and GCMC indicates the data obtained from GCMC simulations under mixture conditions.

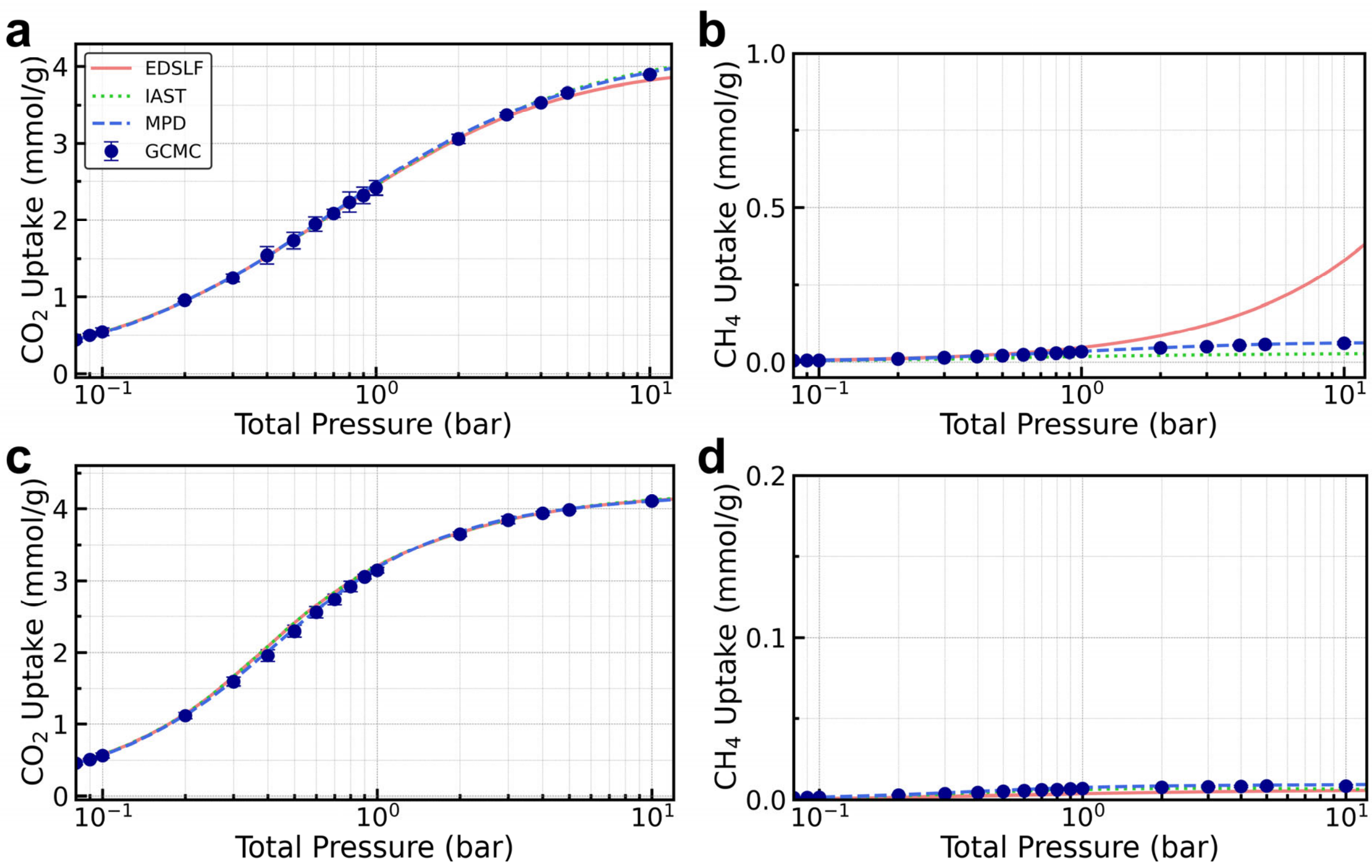

**Figure S4.** Mixture adsorption isotherms of $CO_2$ (left), and $CH_4$ (right) for (a-b) AFG-1 and (c-d) GIS-1 at 298 K with 90/10% $CO_2/CH_4$ mixture. EDSLF indicates the data predicted by EDSLF model, IAST indicates the data predicted by IAST method, MPD indicates the data predicted by MPD-based approach, and GCMC indicates the data obtained from GCMC simulations under mixture conditions.

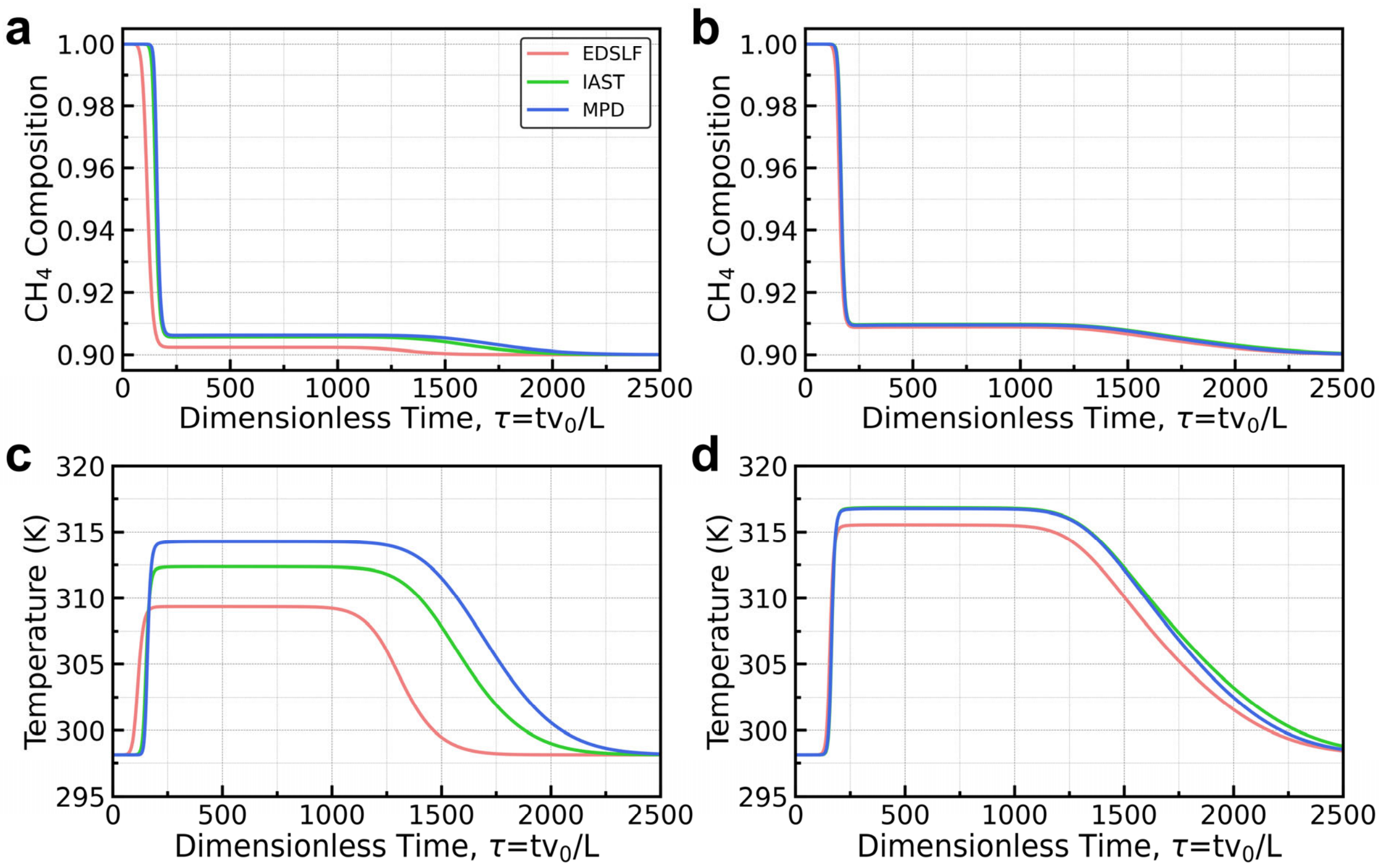

**Figure S5.** $CH_4$ composition breakthrough curves on (a) AFG-1 and (b) GIS-1, and temperature breakthrough curves on (c) AFG-1 and (d) GIS-1 at the exit of the column with 10/90% $CO_2$/$CH_4$ feed mixture at 2 bar and 298 K.

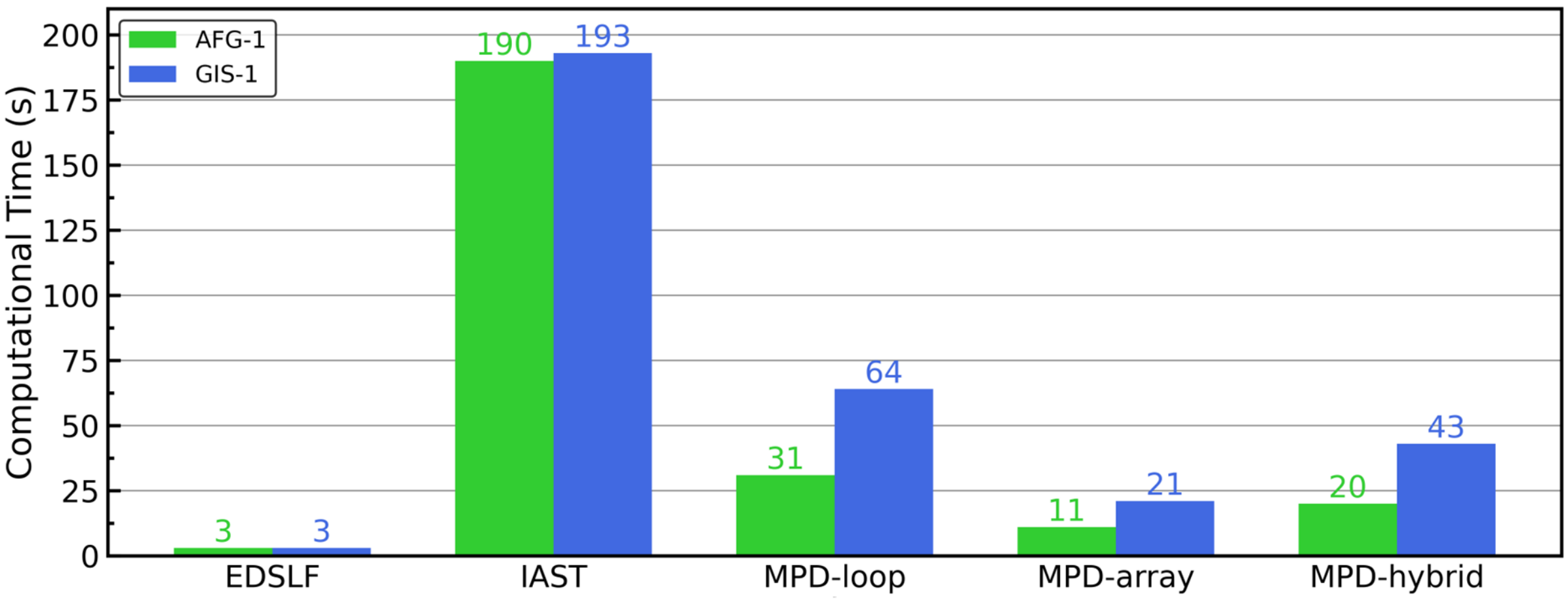

**Figure S6**. Comparison of computational times for EDSLF-, IAST-, and MPD-based breakthrough simulations for binary mixture ($CO_2$/$CH_4$) separation on AFG-1 and GIS-1.

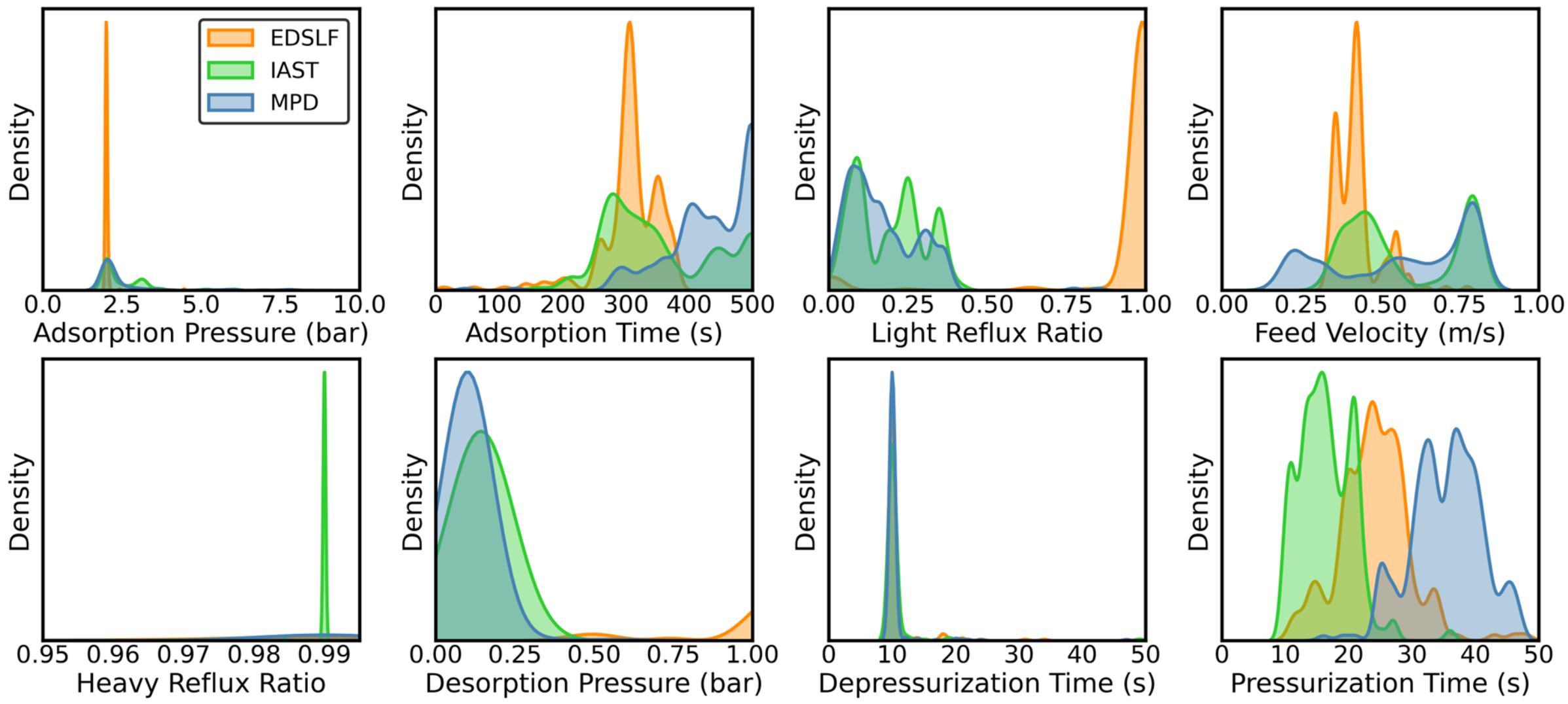

**Figure S7.** Comparison of distributions of optimal decision variables corresponding to Pareto fronts obtained from EDSLF-based, IAST-based, and MPD-based process optimizations for AFG-1.

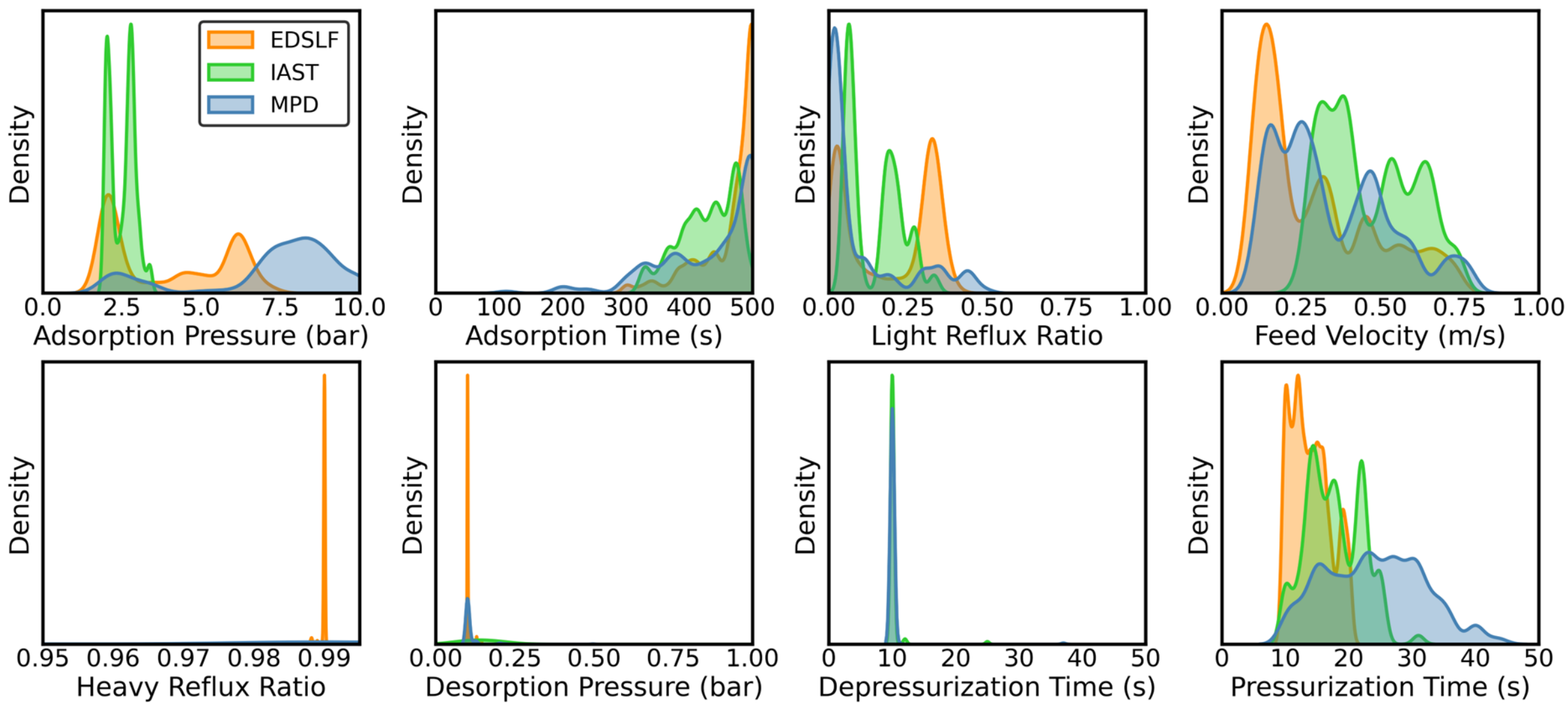

**Figure S8.** Comparison of distributions of optimal decision variables corresponding to Pareto fronts obtained from EDSLF-based, IAST-based, and MPD-based process optimizations for GIS-1.

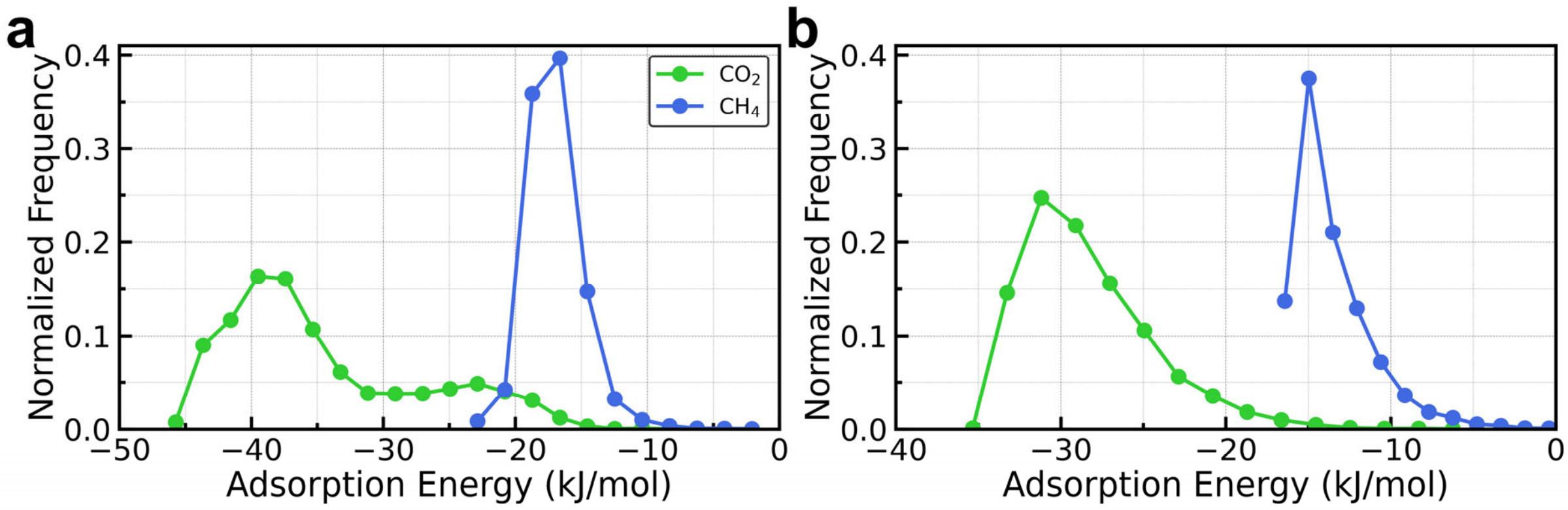

**Figure S9.** Adsorption energy distribution of a single $CO_2$ or $CH_4$ molecule inside (b) LIO-1 and (c) APC-0 at 298 K.

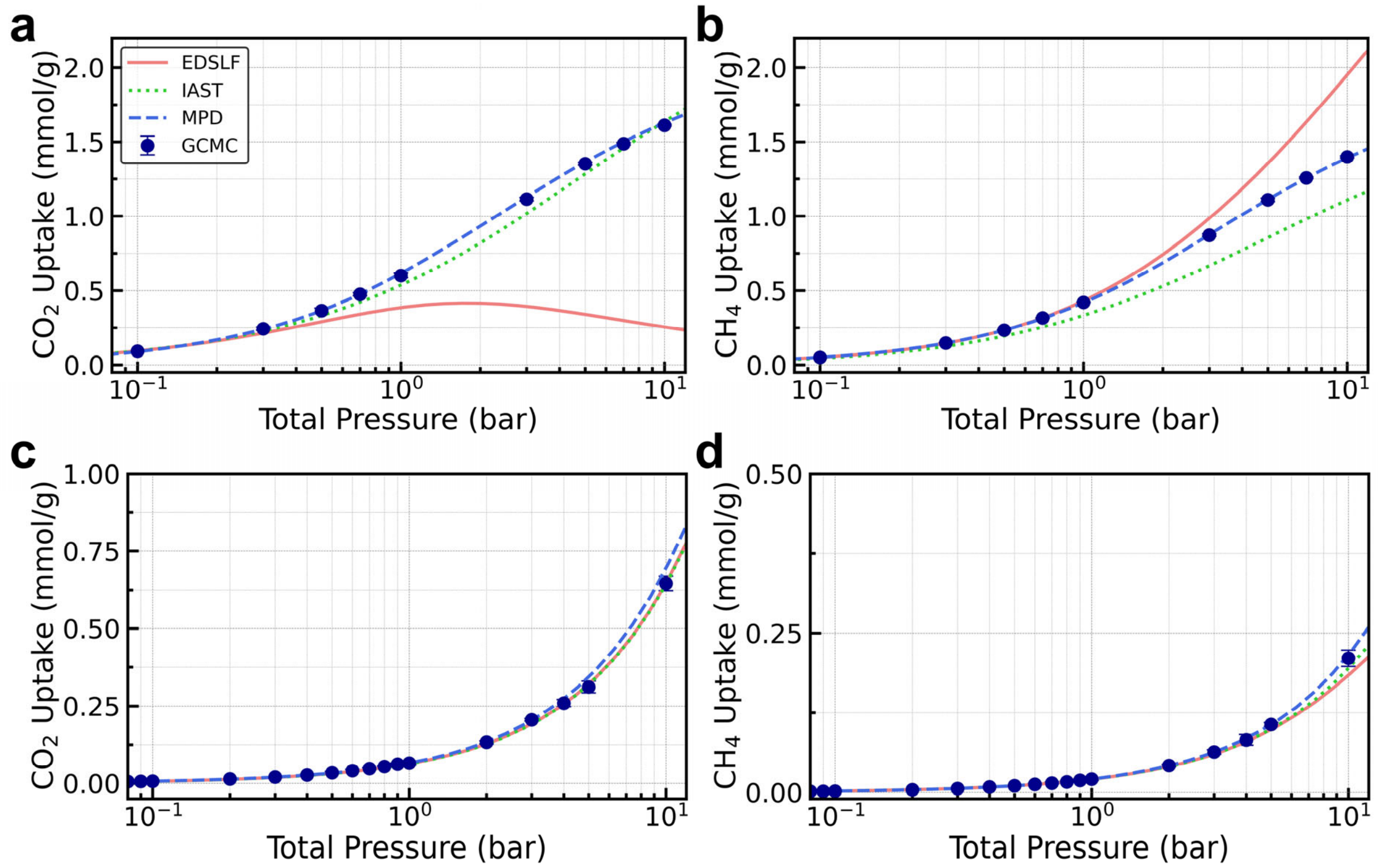

**Figure S10.** Mixture adsorption isotherms of $CO_2$ (left), and $CH_4$ (right) for (a-b) LIO-1 and (c-d) APC-0 at 298 K with 10/90% $CO_2$/$CH_4$ mixture. EDSLF indicates the data predicted by EDSLF model, IAST indicates the data predicted by IAST method, MPD indicates the data predicted by MPD-based approach, and GCMC indicates the data obtained from GCMC simulations under mixture conditions.

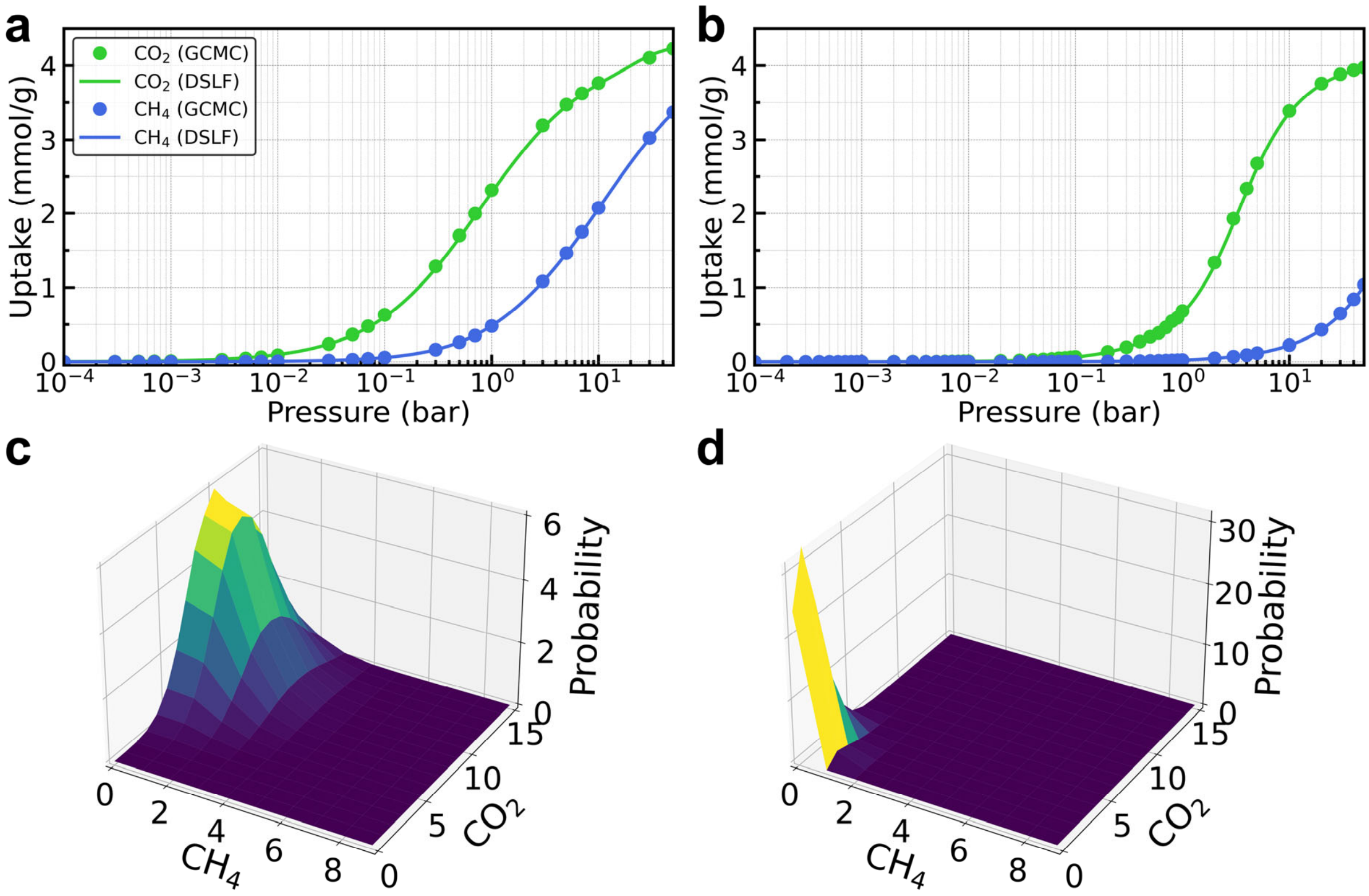

**Figure S11.** Single component adsorption isotherms and fitted models of (a) LIO-1 and (b) APC-0 for $CO_2$ and $CH_4$ at 298 K. GCMC indicates the data obtained from GCMC simulations, and DSLF indicates the data predicted by the DSLF model. 2D MPD (%) for (c) LIO-1 and (d) APC-0 determined by the 2D NVT+W method at reference condition (0.2 bar, 300 K, and 1:1 molar ratio). Purple and yellow colors represent low and high probabilities, respectively.

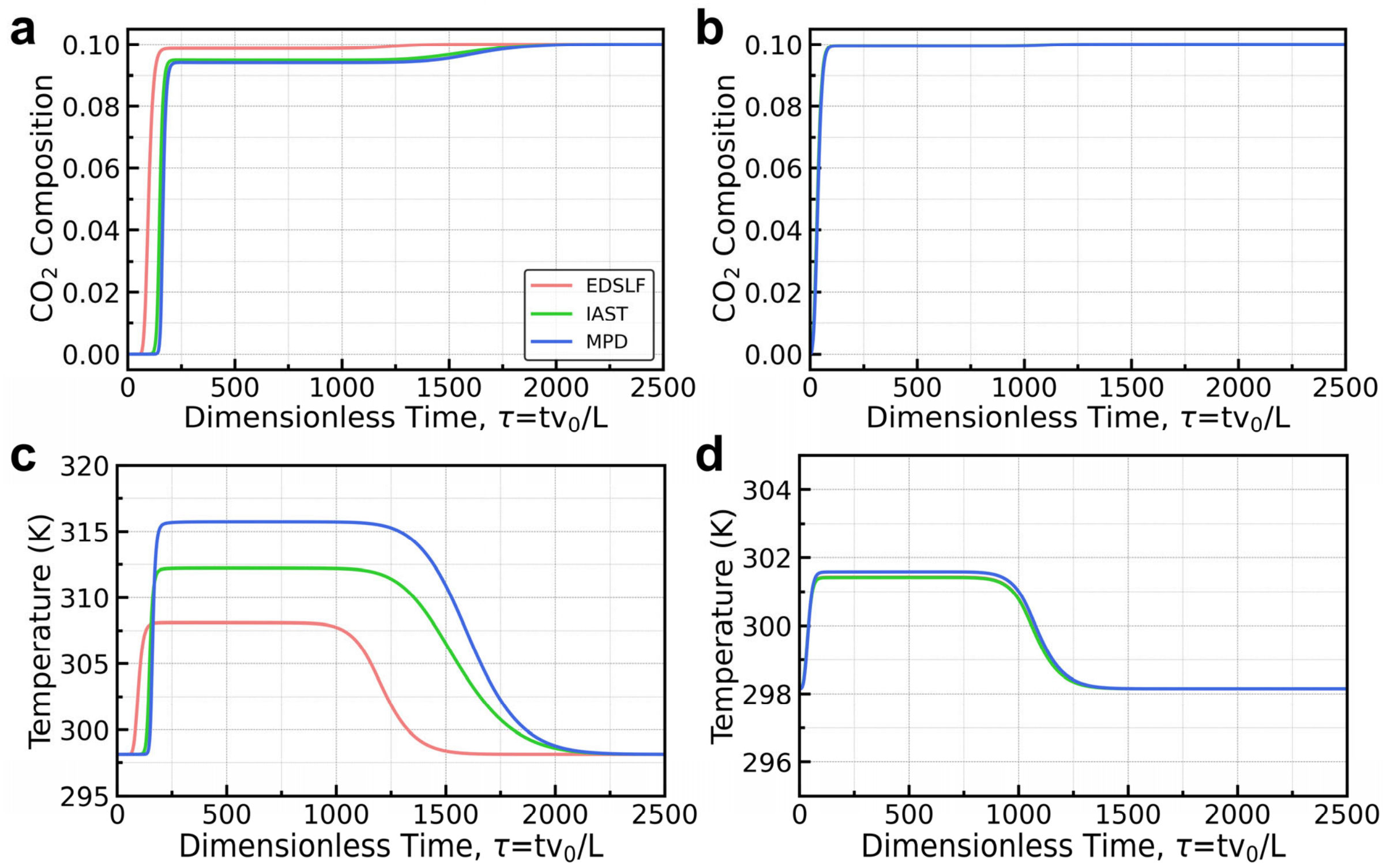

**Figure S12.** $CO_2$ composition breakthrough curves on (a) LIO-1 and (b) APC-0, and temperature breakthrough curves on (c) LIO-1 and (d) APC-0 at the exit of the column with 10/90% $CO_2$/$CH_4$ feed mixture at 2 bar and 298 K.

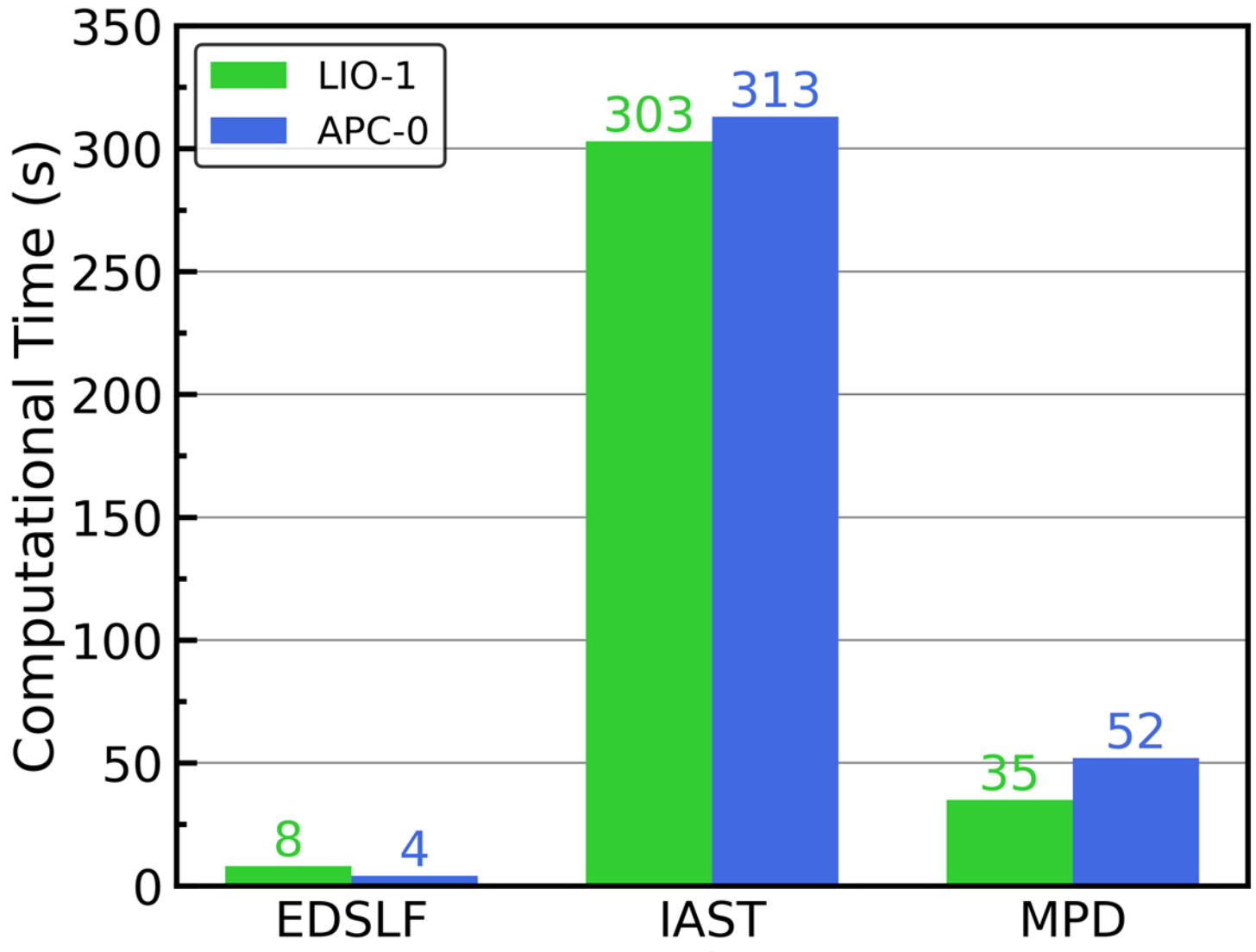

**Figure S13.** Comparison of computational times for EDSLF-, IAST-, and MPD-based breakthrough simulations for binary mixture ($CO_2$/$CH_4$) separation on LIO-1 and APC-0.

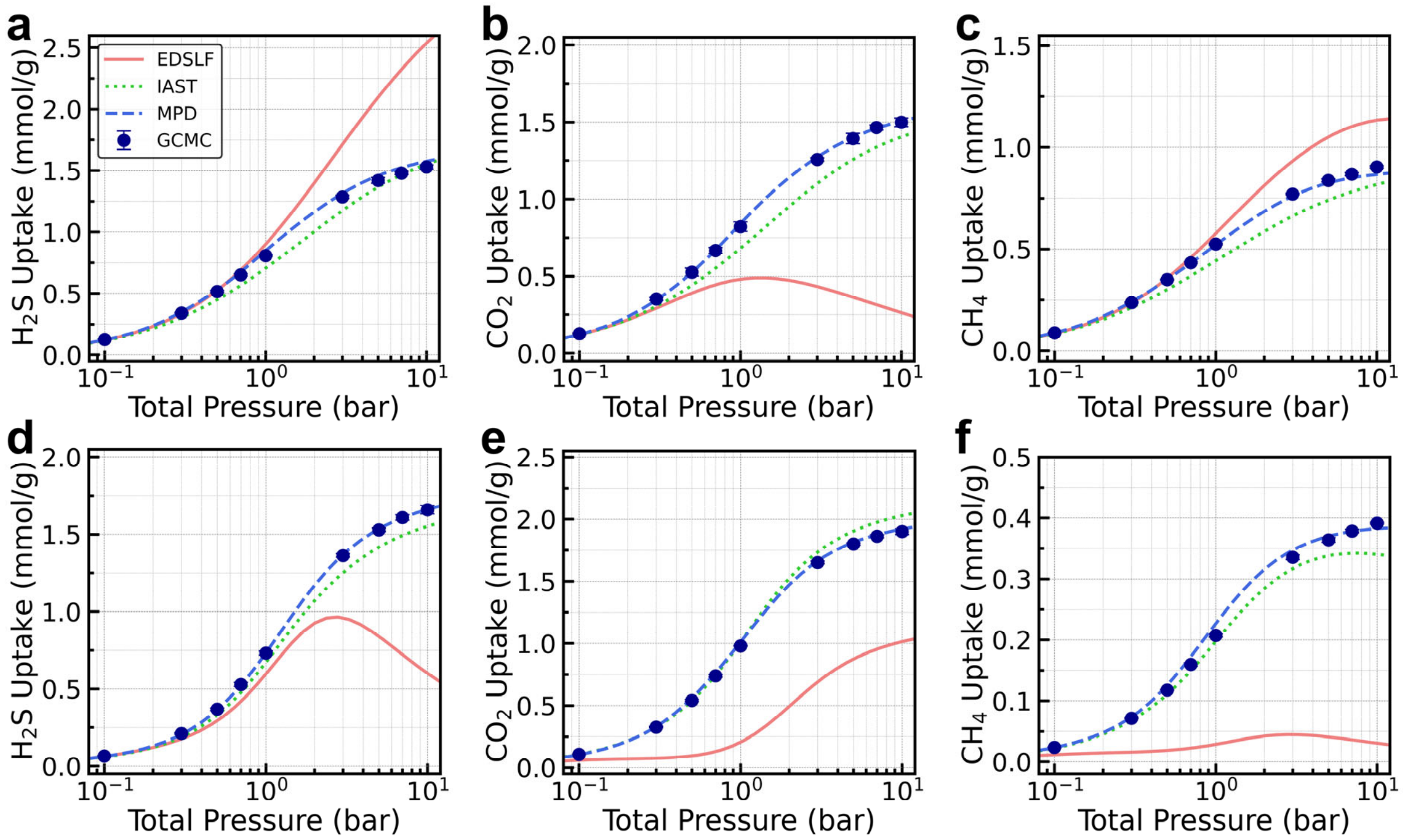

**Figure S14**. Mixture adsorption isotherms of $H_2S$ (left), $CO_2$ (middle), and $CH_4$ (right) for (a-c) AFG-1 and (d-f) GIS-1 at 273 K with 5/5/90% $H_2S/CO_2/CH_4$ mixture. EDSLF indicates the data predicted by EDSLF model, IAST indicates the data predicted by IAST method, MPD indicates the data predicted by MPD-based approach, and GCMC indicates the data obtained from GCMC simulations under mixture conditions.

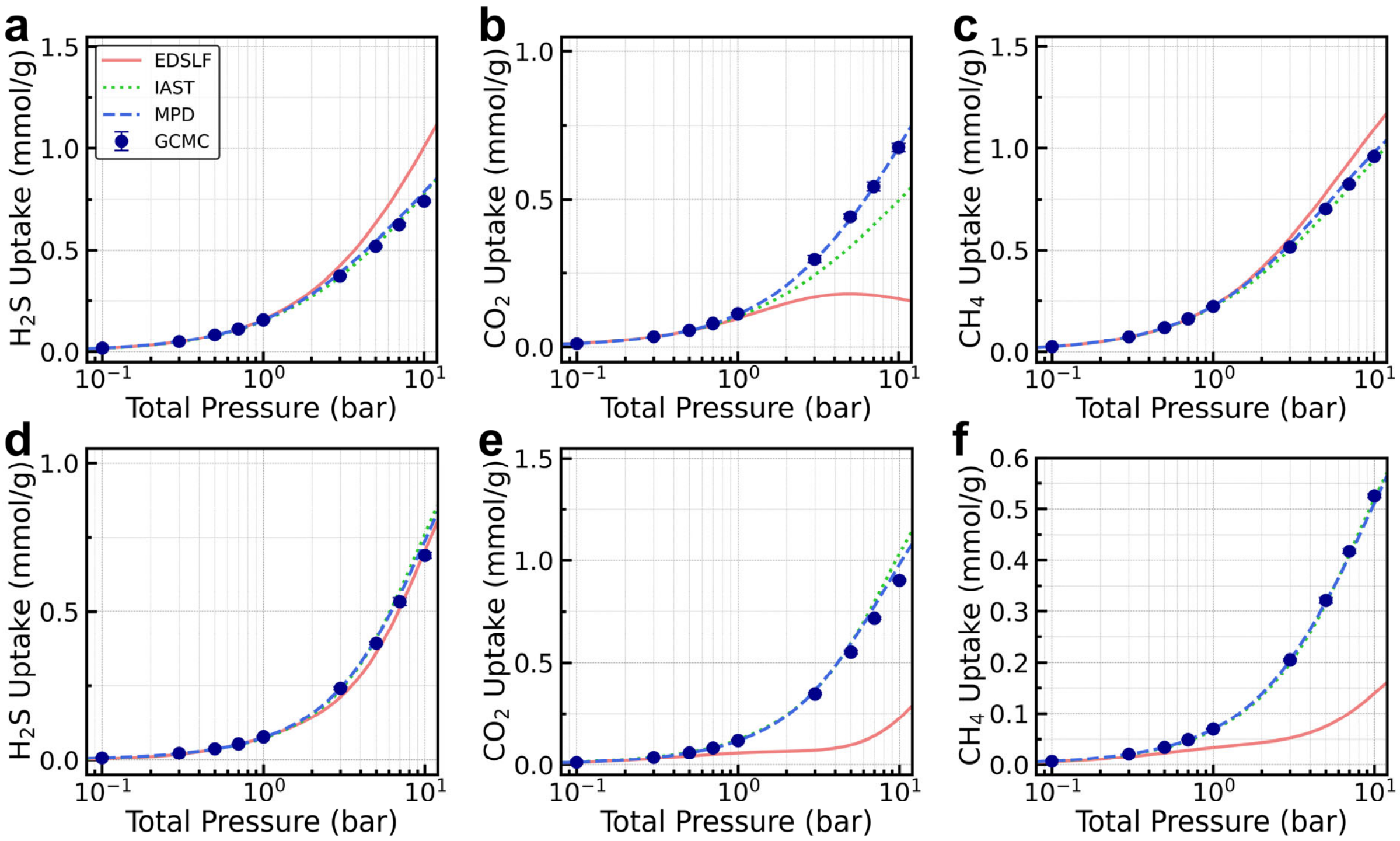

**Figure S15**. Mixture adsorption isotherms of $H_2S$ (left), $CO_2$ (middle), and $CH_4$ (right) for (a-c) AFG-1 and (d-f) GIS-1 at 323 K with 5/5/90% $H_2S/CO_2/CH_4$ mixture. EDSLF indicates the data predicted by EDSLF model, IAST indicates the data predicted by IAST method, MPD indicates the data predicted by MPD-based approach, and GCMC indicates the data obtained from GCMC simulations under mixture conditions.

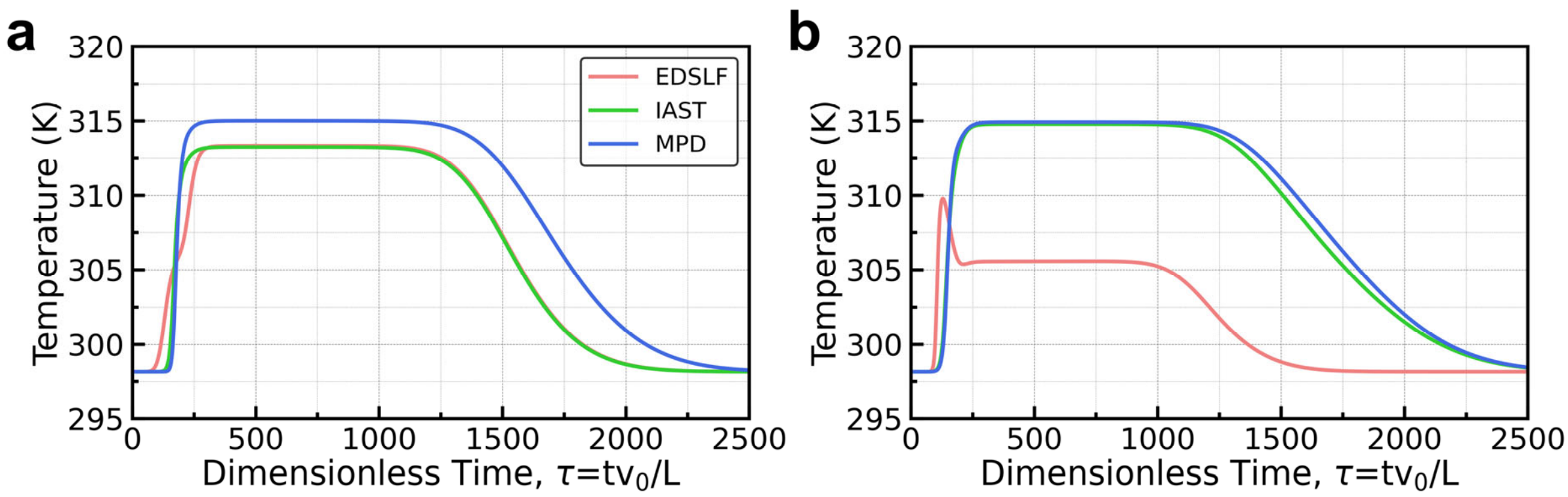

**Figure S16.** Temperature breakthrough curves on (a) AFG-1 and (b) GIS-1 at the exit of the column with 5/5/90% $H_2S/CO_2/CH_4$ feed mixture at 2 bar and 298 K.

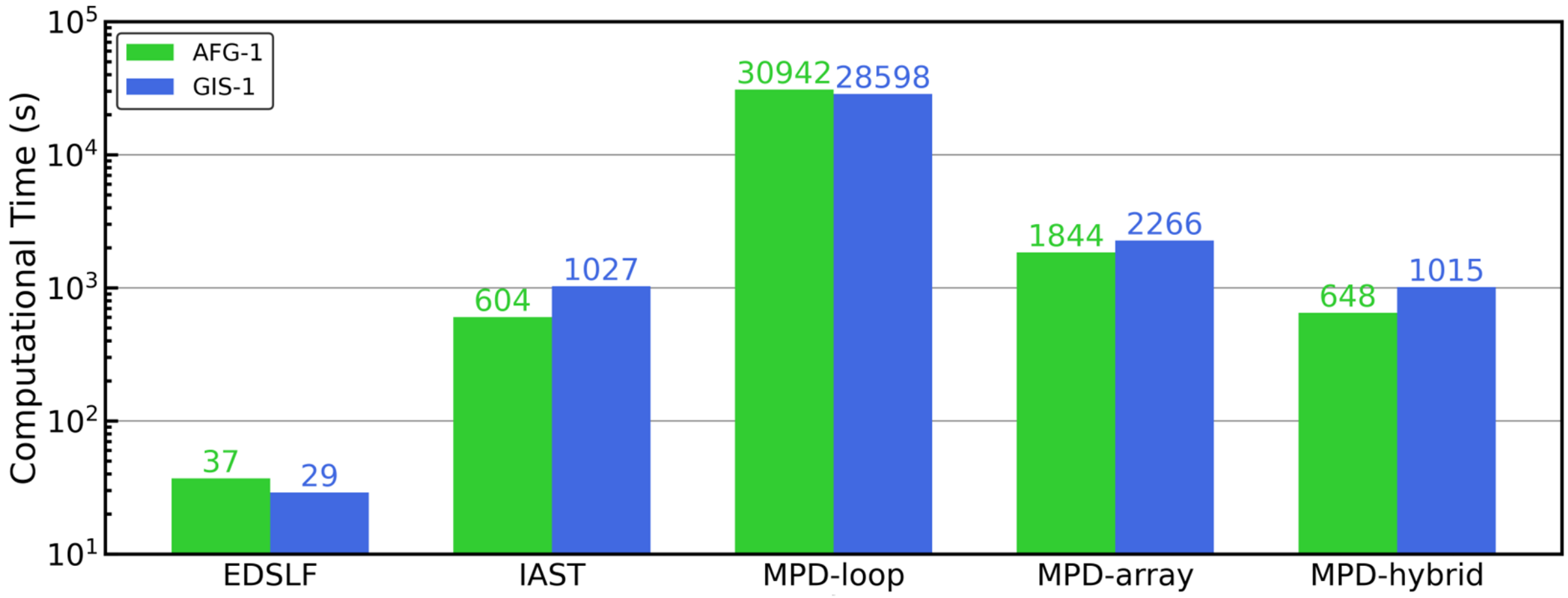

**Figure S17**. Comparison of computational times for EDSLF-, IAST-, and MPD-based breakthrough simulations for ternary mixture ($H_2S$/$CO_2$/$CH_4$) separation on AFG-1 and GIS-1.

## 3. Supplementary tables

**Table S1. All force field parameters used in this work.**

| Atom type | $\varepsilon/k_b(K)$ | $\delta$ (Å) | $q$ $(e)$ |
|---|---|---|---|
| O (Framework) | 53.0 | 3.30 | −0.75 |
| Si (Framework) | 22.0 | 2.30 | 1.50 |
| Al (Framework) | 22.0 | 2.30 | 1.50 |
| P (Framework) | 22.0 | 2.30 | 1.50 |
| O ($CO_2$) | 79.0 | 3.05 | −0.35 |
| C ($CO_2$) | 27.0 | 2.80 | 0.70 |
| S ($H_2S$) | 122.0 | 3.60 | 0.00 |
| H ($H_2S$) | 50.0 | 2.50 | 0.21 |
| M ($H_2S$) | 0.0 | 0.00 | −0.42 |
| $CH_4$ (sp3) | 148.0 | 3.73 | 0.00 |

**Table S2**. **Mathematical model for the pressure/vacuum swing adsorption (PVSA) cycle.**

**Component mass balance:**

$$\frac{\partial y_i}{\partial \tau} = \frac{D_L}{v_0 L}\left(\frac{\partial^2 y_i}{\partial Z^2} + \frac{1}{\bar{P}}\frac{\partial \bar{P}}{\partial Z}\frac{\partial y_i}{\partial Z} - \frac{1}{\bar{T}}\frac{\partial \bar{T}}{\partial Z}\frac{\partial y_i}{\partial Z}\right) - \bar{v}\frac{\partial y_i}{\partial Z} + \frac{(1-\varepsilon)}{\varepsilon}\frac{RT_0 q_{s,0}}{P_0}\frac{\bar{T}}{\bar{P}}\left((y_i - 1)\frac{\partial x_i}{\partial \tau} + y_i \sum_{i,i\neq j}^{n_{comp}} \frac{\partial x_i}{\partial \tau}\right)$$

**Total mass balance:**

$$\frac{1}{\bar{P}}\frac{\partial \bar{P}}{\partial \tau} - \frac{1}{\bar{T}}\frac{\partial \bar{T}}{\partial \tau} = -\frac{\bar{T}}{\bar{P}}\frac{\partial}{\partial Z}\left(\bar{v}\frac{\bar{P}}{\bar{T}}\right) - \frac{(1-\varepsilon)}{\varepsilon}\frac{RT_0 q_{s,0}}{P_0}\frac{\bar{T}}{\bar{P}}\sum_{i=1}^{n_{comp}}\frac{\partial x_i}{\partial \tau}$$

**Solid phase mass balance:**

$$\frac{\partial x_i}{\partial \tau} = \frac{k_i L}{v_0}(x_i^* - x_i)$$

**Column energy balance:**

$$\left(\varepsilon C_g C_{p,g} + (1-\varepsilon)\left(C_{p,a} q_{s,0} + C_{p,s}\rho_s\right)\right)\frac{\partial \bar{T}}{\partial \tau} = \frac{K_z}{v_0 L}\frac{\partial^2 \bar{T}}{\partial Z^2} - \bar{v}\varepsilon C_g C_{p,g}\frac{\partial \bar{T}}{\partial Z} - (1-\varepsilon)\frac{q_{s,0}}{T_0}\sum_{i=1}^{n_{comp}}(\Delta H_i)\frac{\partial x_i}{\partial \tau}$$

**Pressure drop:**

$$-\frac{\partial \bar{P}}{\partial Z} = \frac{150\mu(1-\varepsilon)^2 L v_0}{4r_p^2\varepsilon^2 P_0}\bar{v} + \frac{1.75 L v_0^2}{2r_p T_0}\frac{(1-\varepsilon)}{\varepsilon}\left(\sum_i MW_i \frac{y_i \bar{P}}{R\bar{T}}\right)\bar{v}\|\bar{v}\|$$

**Dimensionless variables:**

$$\bar{P} = \frac{P}{P_0};\ \bar{T} = \frac{T}{T_0};\ x_i = \frac{q_i\rho_s}{q_{s,0}};\ x_i^* = \frac{q_i^*\rho_s}{q_{s,0}};\ \bar{v} = \frac{v_z}{v_0};\ Z = \frac{z}{L};\ \tau = \frac{t v_0}{L}$$

**Table S3**. **Boundary conditions for each step of the modified Skarstrom cycle.**

| | Pressure ($\bar{P}$) | Temperature ($\bar{T}$) | Mole fraction ($y_i$) |
|---|---|---|---|
| | *At the entrance of the column (Z=0)* | | |
| Pressurization | $\bar{P} = \overline{P_L} \rightarrow 1$ | $y_i = y_{i,feed}$ | $\bar{T} = 1$ |
| Adsorption | $\bar{P} = 1.02$ | $y_i = y_{i,feed}$ | $\bar{T} = 1$ |
| Heavy reflux | $\bar{P} = 1.02$ | $y_i = y_{i,LR}\|_{Z=0}$ | $\bar{T} = \overline{T_{LR}}\|_{Z=0}$ |
| Depressurization | $\bar{P} = 1 \rightarrow \overline{P_L}$ | $\frac{\partial y_i}{\partial Z} = 0$ | $\frac{\partial \bar{T}}{\partial Z} = 0$ |
| Light reflux | $\bar{P} = \overline{P_L}$ | $\frac{\partial y_i}{\partial Z} = 0$ | $\frac{\partial \bar{T}}{\partial Z} = 0$ |
| | *At the end of the column (Z=1)* | | |
| Pressurization | $\frac{\partial \bar{P}}{\partial Z} = 0$ | $\frac{\partial y_i}{\partial Z} = 0$ | $\frac{\partial \bar{T}}{\partial Z} = 0$ |
| Adsorption | $\bar{P} = 1$ | $\frac{\partial y_i}{\partial Z} = 0$ | $\frac{\partial \bar{T}}{\partial Z} = 0$ |
| Heavy reflux | $\bar{P} = 1$ | $\frac{\partial y_i}{\partial Z} = 0$ | $\frac{\partial \bar{T}}{\partial Z} = 0$ |
| Depressurization | $\frac{\partial \bar{P}}{\partial Z} = 0$ | $\frac{\partial y_i}{\partial Z} = 0$ | $\frac{\partial \bar{T}}{\partial Z} = 0$ |
| Light reflux | $\bar{P} > \overline{P_L}$ | $y_i = y_{i,Ads}\|_{Z=1}$ | $\bar{T} = \overline{T_{Ads}}\|_{Z=1}$ |

**Table S4**. **DSLF parameters with the corresponding $R^2$ values for AFG-1 and GIS-1.**

| Parameter | AFG-1 | | | GIS-1 | | |
|---|---|---|---|---|---|---|
| | $H_2S$ | $CO_2$ | $CH_4$ | $H_2S$ | $CO_2$ | $CH_4$ |
| $q_{sb,i}$ | 4.11 | 7.03 | 4.16 | 0.11 | 3.08 | 1.10 |
| $q_{sd,i}$ | 0.16 | 4.21 | 0.35 | 4.03 | 1.17 | 2.54 |
| $b_i$ | 1.78e−5 | 3.45e−7 | 2.84e−6 | 1.63e−5 | 1.75e−5 | 5.24e−7 |
| $d_i$ | 7.85e−16 | 1.94e−5 | 2.38e−8 | 2.22e−8 | 9.81e−12 | 4.44e−12 |
| $n_{b,i}$ | 0.98 | 1.38 | 1.12 | 0.68 | 0.97 | 0.93 |
| $n_{d,i}$ | 0.33 | 1.02 | 0.68 | 0.61 | 0.41 | 0.56 |
| $R^2$ | 1.00 | 1.00 | 1.00 | 1.00 | 1.00 | 1.00 |

**Table S5**. **Heat of adsorption ($\Delta H_i$) of $H_2S$, $CO_2$, and $CH_4$ in AFG-1 and GIS-1 at 298 K.**

| Zeolite | Heat of adsorption, $\Delta H_i$ (kJ/mol) | | |
|---|---|---|---|
| | $H_2S$ | $CO_2$ | $CH_4$ |
| AFG-1 | −29.29 | −35.22 | −19.02 |
| GIS-1 | −30.80 | −31.95 | −17.75 |

**Table S6**. **DSLF parameters with the corresponding $R^2$ values for LIO-1 and APC-0.**

| Parameter | LIO-1 | | APC-0 | |
|---|---|---|---|---|
| | $CO_2$ | $CH_4$ | $CO_2$ | $CH_4$ |
| $q_{sb,i}$ | 0.19 | 3.93 | 2.36 | 0.00 |
| $q_{sd,i}$ | 4.17 | 0.26 | 1.76 | 15.36 |
| $b_i$ | 1.25e−22 | 2.68e−6 | 2.24e−6 | 0.00 |
| $d_i$ | 6.42e−5 | 5.45e−8 | 2.96e−12 | 1.66e−8 |
| $n_{b,i}$ | 0.29 | 1.09 | 0.98 | 1.00 |
| $n_{d,i}$ | 1.17 | 0.70 | 0.48 | 1.01 |
| $R^2$ | 1.00 | 1.00 | 1.00 | 1.00 |

**Table S7**. **Heat of adsorption ($\Delta H_i$) of $CO_2$ and $CH_4$ in LIO-1 and APC-0 at 298 K.**

| Zeolite | Heat of adsorption, $\Delta H_i$ (kJ/mol) | |
|---|---|---|
| | $CO_2$ | $CH_4$ |
| LIO-1 | −37.10 | −18.82 |
| APC-0 | −30.06 | −15.67 |